\documentclass[twocolumn]{aastex7}

\usepackage{natbib}
\usepackage{multirow}
\usepackage{booktabs}
\usepackage{epsfig}
\usepackage{ulem}
\usepackage{hyperref}
\usepackage{rotating}
\usepackage{graphicx}
\usepackage{url}
\usepackage{amsmath}
\usepackage{color}
\usepackage{savesym}
\savesymbol{tablenum}
\usepackage{siunitx}
\usepackage[normalem]{ulem}
\usepackage{enumerate}
\usepackage{gensymb}
\usepackage{xcolor}
\usepackage{xspace}
\usepackage{longtable}
\usepackage{afterpage}
\usepackage{threeparttable}
\usepackage{tabularx}
\usepackage[bottom]{footmisc}

\newcommand\rev[1]{{\color{black}#1}}
\newcommand\revtwo[1]{{\color{black}#1}}
\newcommand\revthree[1]{{\color{black}#1}}
\DeclareSIUnit\parsec{pc}
\DeclareSIUnit\h{\textit{h}}

\graphicspath{{./}{figs/}}

\DeclareRobustCommand{\uvec}[1]{{%
  \ifcsname uvec#1\endcsname
     \csname uvec#1\endcsname
   \else
    \bm{\hat{\mathbf{#1}}}%
   \fi
}}

\begin{document}
\title{TITAN DR1: An Improved, Validated, and Systematically-Controlled Recalibration of \\ ATLAS Photometry toward Type Ia Supernova Cosmology}
\shorttitle{TITAN DR1: Systematics Controlled Calibration}
\shortauthors{Marlin et al.}

\correspondingauthor{Elijah G. Marlin, Yukei S. Murakami}

\author[0009-0003-4631-3184]{Elijah G. Marlin}
\email[show]{emarlin@bu.edu}
\affiliation{Departments of Astronomy and Physics, Boston University, Boston MA 02215}

\author[0000-0002-8342-3804]{Yukei S. Murakami}
\email[show]{ymuraka2@jhu.edu}
\affiliation{Department of Physics and Astronomy, Johns Hopkins University, Baltimore, MD 21218, USA}

\author[0000-0001-5201-8374]{Dillon Brout}
\email{}
\affiliation{Departments of Astronomy and Physics, Boston University, Boston MA 02215}

\author[0009-0004-5681-545X]{Jack W. Tweddle}
\email{jack.tweddle@physics.ox.ac.uk}
\affiliation{Astrophysics sub-Department, Department of Physics, University of Oxford, Keble Road, Oxford, OX1 3RH, UK}

\author[0000-0002-8012-6978]{Brodie Popovic}
\email{}
\affiliation{School of Physics and Astronomy, University of Southampton, Southampton, UK, SO17 1BJ}

\author[0000-0001-9535-3199]{Ken W. Smith}
\email{}
\affiliation{Astrophysics sub-Department, Department of Physics, University of Oxford, Keble Road, Oxford, OX1 3RH}
\affiliation{Astrophysics Research Centre, School of Mathematics and Physics, Queen’s University Belfast, BT7 1NN, UK}

\author[0000-0002-8229-1731]{Stephen J. Smartt}
\email{}
\affiliation{Astrophysics sub-Department, Department of Physics, University of Oxford, Keble Road, Oxford, OX1 3RH, UK}
\affiliation{Astrophysics Research Centre, School of Mathematics and Physics, Queen’s University Belfast, BT7 1NN, UK}

\author[0000-0002-4934-5849]{Daniel M. Scolnic}
\email{}
\affiliation{Department of Physics, Duke University, Durham, NC 27708, USA}

\author[0000-0002-6230-0151]{David Jones}
\email{}
\affiliation{Institute for Astronomy, University of Hawai'i, 640 N.~A'ohoku Pl., Hilo, HI 96720, USA}

\author[0000-0001-8596-4746]{Erik R. Peterson}
\email{}
\affiliation{Department of Physics, University of Michigan, Ann Arbor, MI 48109, USA}
\affiliation{Society of Fellows, University of Michigan, Ann Arbor, MI 48109, USA}

\author[0000-0002-6124-1196]{Adam G. Riess}
\email{ariess@stsci.edu}
\affiliation{Department of Physics and Astronomy, Johns Hopkins University, Baltimore, MD 21218, USA}
\affiliation{Space Telescope Science Institute, Baltimore, MD 21218, USA}

\author[0000-0001-8788-1688]{Maria Vincenzi}
\email{}
\affiliation{Astrophysics sub-Department, Department of Physics, University of Oxford, Keble Road, Oxford, OX1 3RH, UK}

\author[0000-0001-5399-0114]{Nora F. Sherman}
\email{}
\affiliation{Institute for Astrophysical Research, Department of Astronomy, Boston University, 725 Commonwealth Avenue, Boston, MA 02215, USA}

\author[0000-0001-5399-0114]{Maria Acevedo}
\email{}
\affiliation{Department of Physics, Duke University, Durham, NC 27708, USA}

\author[0009-0002-7557-0406]{Jasper Milstein}
\email{}
\affiliation{Departments of Astronomy and Physics, Boston University, Boston MA 02215}

\author[0000-0003-0928-0494]{Mitchell Dixon}
\email{}
\affiliation{Institute for Astronomy, University of Hawai'i, 640 N.~A'ohoku Pl., Hilo, HI 96720, USA}

\author[0000-0002-4410-5387]{Armin Rest}
\email{}
\affiliation{Department of Physics and Astronomy, Johns Hopkins University, Baltimore, MD 21218, USA}
\affiliation{Space Telescope Science Institute, Baltimore, MD 21218, USA}

\setcounter{footnote}{0}

\begin{abstract}
ATLAS (Asteroid Terrestrial Last Alert System) is a time-domain survey using four telescopes, covering the entire sky. It has observed over 10,000 spectroscopically confirmed Type Ia supernovae (SNe~Ia), with thousands of cosmology-grade light curves (to be released as TITAN DR1). To prepare this massive, low-redshift dataset for cosmology, we evaluate and cross-calibrate ATLAS forced photometry using tertiary stars from the DES (Dark Energy Survey) Y6 release. The 5000 deg$^2$ DES footprint overlaps regions both in and out of the PS1 (Pan-STARRS DR1) footprint, allowing tests of the primary calibrator for the ATLAS Refcat2 catalog. Initial offsets are at the $\sim$40 mmag scale. To improve this we determine $\Delta$ zeropoint offsets for two cases: (1) pixel-to-pixel offsets within individual CCDs (reduced from $\sim$8 to $\sim$4 mmag RMS) and (2) chip-to-chip offsets across the 9 CCDs and filters (reduced from $\sim$17 to $\sim$3 mmag RMS). We also identify the largest systematic uncertainty as a transmission-function color dependence, requiring shifts in the assumed ATLAS filters at the $\sim$30 mmag level if uncorrected. We validate our calibration using (a) CALSPEC standards, (b) an independent tertiary catalog, and (c) distance moduli of cross-matched SNe~Ia, all showing improved consistency. Overall, we estimate combined calibration-related systematics at the $\sim$5--10 mmag level, supporting competitive cosmological constraints with the TITAN SN~Ia dataset.

\end{abstract}


\section{Introduction}
\setcounter{footnote}{0}

Type Ia supernovae (SNe~Ia), thermonuclear explosions of  white dwarf stars, are one of the most successful standardizable candles thanks to their known luminosity-color-duration relationship \citep{Phillips93, Hamuy96, Tripp98}. 
The small scatter in the post-standardization luminosity makes SNe~Ia an excellent distance indicator for cosmology \citep[e.g.,][]{Filippenko_05_cosmology}. The state-of-the-art measurements of cosmological parameters, including the equation of state for dark energy \citep[e.g.,][]{ DESI_2025_BAO}, use a compilation of SNe~Ia samples that cover a wide range of redshifts, such as DESY5 \citep{sanchez2024darkenergysurveysupernova}, Pantheon+ \citep{Brout_2022,Scolnic_2022_Pantheon+}, and UNION3 \citep{Rubin_2025_UNION}.

A commonality among these datasets is that they combine low-redshift ($z\lesssim0.1$) and high-redshift ($z\lesssim1$) surveys. All SNe~Ia datasets used in \cite{DESI_2025_BAO}, DES  (1500 high-$z$ SNe~Ia), UNION3 (containing more than 2000 high-$z$ SNe~Ia) and Pantheon+ (1550 high-$z$ SNe~Ia) take advantage of \rev{a common set of historical} low-$z$ datasets which add up to $\sim$200 SNe~Ia, (e.g., CfA1; \citealt{Riess99}, CfA2; \citealt{Jha06}, CfA3-Keplercam; \citealt{Hicken09a}, CfA3-4Shooter; \citealt{Hicken09b}, CfA4p1, CfA4p2; \citealt{Hicken12}, CSP DR3; \citealt{Krisciunas17}, LOSS1; \citealt{Ganeshalingam_2010_LowZ}, LOSS2; \citealt{Stahl19}, SOUSA; \citealt{Brown_2015_LowZ}, Foundation; \citealt{Foley18}, CNIa0.02; \citealt{Chen_2022_LowZ}). Current constraints on cosmology rely on these historical low-$z$ SN Ia datasets to point to interesting new physics \citep[e.g.]{Boruah20a,Riess22,Brout22,vincenzi2024,Abbott_2024_DES,DESI_2025_BAO,tang2025unitingobserveddynamicaldark}. This reliance on existing low-$z$ datasets is problematic, 1) because the number of low-$z$ SNe~Ia \rev{remains \revtwo{relatively} constant whereas high-$z$ datasets are growing rapidly with dedicated surveys}, and 2) \rev{because all analyses rely on this set of supernovae used not only for constraining nearby distances but also for training the underlying SN~Ia model, this means that all cosmological analyses are inherently correlated.} These will continue to be challenges for harnessing the full potential of upcoming flagship high-$z$ surveys, such as LSST \citep{Foley_2018_LSST_proposal_1} and NASA Roman \citep{Sanderson_2024_Roman}. This necessitates a renewed focus on the collection and analysis of precision low-$z$ datasets, collected over many years.

\rev{While high-$z$ SN~Ia samples are expanding rapidly, \revtwo{the collection of a larger and less biased low-$z$ SNe~Ia poses a challenge.} The volumetric SN~Ia rate of $\sim2\times10^{-5}\,\mathrm{SNe~Ia~yr^{-1}~Mpc^{-3}}$ corresponds to roughly one SN~Ia per galaxy per century \citep{Dilday_2010_low-z}. Even surveying the full sky, this yields only an order of hundreds of SNe~Ia per year within $z\sim0.1$. Consequently, building a high-quality, spectroscopically confirmed low-$z$ sample is inherently time--limited and requires continuous, nearly all-sky monitoring over many years. Upcoming surveys such as LSST will discover vast numbers of high-$z$ SNe~Ia, but LSST cannot rapidly discover low-$z$ SNe~Ia, since the limited nearby volume fixes the pace at which new low-$z$ SNe~Ia appear. In contrast to the explosive growth of high-$z$ datasets, the buildup of precision low-$z$ samples will remain a slow, volume-limited endeavor.} \revtwo{In recent years, several new low-$z$ SN~Ia surveys have begun to expand this nearby sample, including the Zwicky Transient Facility (ZTF; \citealt{Rigault_25_ZTFRelease}), the Young Supernova Experiment (YSE; \citealt{Aleo_23_YSERelease}), and DEBASS \citep{Sherman_25_DEBASSRelease,acevedo2025darkenergybedrockallsky}. These programs have made important contributions by increasing discovery rates and providing well-sampled light curves over limited sky areas. Differences in survey strategy, footprint, and calibration approaches mean that no single program yet provides a uniform, all-sky, long-baseline, cosmology-ready low-$z$ ($z<0.1$) SN~Ia dataset of greater than a few hundred.}

The TITAN (The Type Ia supernova Trove from ATLAS in the Nearby universe) SN Ia dataset that we present in this series of papers will provide a solution with several thousand at low-$z$. TITAN is a compilation of spectroscopically-confirmed SNe~Ia observed by the Asteroid Terrestrial-impact Last Alert System (ATLAS;  \citealt{Tonry_2018_atlas}). ATLAS, a NASA-funded all-sky survey, visits the whole sky every night with limited magnitudes at $\revtwo{m}\sim20$ mag, making it optimal for capturing low-$z$ SNe~Ia up to $z\lesssim0.1$. The first data release of the TITAN dataset, containing $\sim10,000$ light curves ($\sim3000$ cosmology grade with host--galaxy $z$) consists of four papers. The overview, SN Ia light curves, and the Hubble diagram are presented in \cite{Murakami_TITAN_DR1}; and  the association of SNe~Ia with their host galaxies, the compilation of redshifts, and the determination of galaxy properties is the subject of \cite{Tweddle_2026_hosts}; and the simulation and the forward--modeling of observational bias is presented in \cite{Tweddle_2026_simulation}. In this paper, we \rev{externally validate the ATLAS calibration, motivate photometric corrections, perform a preliminary calibration systematic assessment, in preparation for a future} cosmological analysis.

The calibration of datasets, such as that performed in \cite{Popovic_2025_Dovekie} (hereafter Dovekie), consist of two steps: characterization of surveys' photometric systems (e.g., uniformity of the focal plane, temporal changes in transmission properties, linearity along wavelength and flux levels) and correction for each filter/ detector configuration to a single reference photometric system. \cite{Scolnic15} (hereafter Supercal) calculate relative zeropoint offsets using CALSPEC standard and a cross-validation with thousands of tertiary stars overlapping PS1 (Pan-STARRS DR1) and other telescope systems. This method was updated and improved upon in \cite{Brout_22_fragalistic} (hereafter Fragilistic) for Pantheon+, \rev{by allowing all surveys cross-calibrated simultaneously without fixing PS1, allowing for the production of a calibration covariance systematic error budget.} Additionally, Fragilistic quantify the small variations of transmission functions and their impact on cosmology. Dovekie is the most recent iteration of this method, providing an open source framework, \rev{an improved likelihood}, and expanded sets of primary calibration stars with faint DA white dwarf stars. For calibration of the TITAN dataset in this work, we employ the same techniques in order to cross-check consistency of the existing ATLAS calibration \rev{with external datasets (HST Calspec, DA white dwarfs, Dark Energy Survey Y6 Wide Field Catalog)}.

ATLAS is a telescope network comprised of four telescopes, two in the Northern Hemisphere in Hawaii (Dec $\ge -50^\circ$), and two in the Southern Hemisphere in South Africa and Chile (Dec $\lesssim +40^\circ$) \citep{Tonry_2018_atlas}\footnote{ATLAS now has a fifth unit in Tenerife (ATLAS-TEIDE) that is operating as part of the survey. This is a different modular design constructed of 16 Celestron RSA 11 telescopes which with a CMOS camera \citep{Licandro_2023_ATLAS-teide}.  We do not use ATLAS-TEIDE data in any of the TITAN papers.}. 
\revtwo{
An ATLAS camera system consists of a physical CCD device 
(all are STA-1600 devices with format $10560\times10560$ pixels)
The northern telescopes underwent several changes to the camera configurations which we document in Tab.~\ref{tab:detector_summary}. In particular, the Mauna Loa unit had 4 camera changes while Haleakala has had the same CCD device throughout the survey. We define nine separate telescope and camera  combinations over the decade since commissioning of the first unit on Haleakala and we label each with a `chip ID'. Each `chip' that we have defined may not be a unique CCD device. Rather, it is some combination of unique cryostat, detector, and controller. For cosmology we treat them each `chip ID` as  independent systems. 
ATLAS primarily uses two broadband filters: ATLAS-cyan ($4200\lesssim\lambda_\mathrm{obs}\lesssim6500 \mathrm{\AA}$) and ATLAS-orange ($5600\lesssim\lambda_\mathrm{obs}\lesssim8200 \mathrm{\AA}$) \citep{Tonry_2018_atlas}. With these two filter,  we have 18 possible filter-camera configurations e.g., chip 0 -- filter cyan, which we refer to as `chip 0$c$'). However for chips 0 and 2, the cyan filter was never used, which resulted in 16 filter-camera combination (see Tab.~\ref{tab:detector_summary}). In this work we treat each combination as separate filters (similarly to CfA filters in Supercal). }

The baseline ATLAS calibration, applied to every \rev{exposure in the default ATLAS data reduction pipeline}, uses Refcat2, an all--sky tertiary star catalog in the PS1 system \citep{Tonry_2018_refcat}. Refcat2 is comprised of photometry from eight distinct stellar surveys, primarily PS1, Gaia DR2, and APASS. The magnitude of each star is calculated as the average magnitude from each survey that observes it, \rev{opening up for potential mmag-level discontinuities across the sky}. In this work, we use an independent, well-calibrated tertiary star catalog, that covers declination ranges inside and outside PS1 to validate the baseline calibration with Refcat2. We select the DES Y6 tertiary star catalog \citep{Rykoff2023} which is known to have a photometric uniformity of $<$1.8 mmag and whose absolute flux is known at the 1\% level, making it an excellent candidate for a relative calibration. The footprint of 5000 square degrees also provides a wide range of stellar photometry (with over 17 million observed stars) facilitating cross comparison, which is needed given the all sky nature of ATLAS.

We present the data used in this work, including ATLAS, DES, and synthetic photometry in Sec.~\ref{sec:data_prep}. In Sec.~\ref{sec:analysis}, we quantify and discuss two levels of calibration (intra--chip, inter--chip) needed to prepare TITAN for cosmology. \rev{In Sec.~\ref{sec:validation_tests} we demonstrate the tests used to show a validation of our calibration.} We compare the resulting SN Ia luminosities with other modern low-$z$ datasets in Sec.~\ref{sec:dinstance_comparison}. We discuss the implications of our findings and their impact on cosmology in Sec.~\ref{sec:discussion}, followed by our concluding results in Sec.~\ref{sec:conclusion}. 

\begin{deluxetable*}{ccccccc}
        \vspace{3mm}
        \tablehead{
        \colhead{chip ID} &\colhead{Site} & \colhead{CCD nickname} & \colhead{cryostat} & \colhead{Serial Number} & \colhead{MJD$_\mathrm{min}$} & \colhead{MJD$_\mathrm{max}$}
        }
        \startdata
        0  &  01a & fuzzy   & gold  & STA1600LN-SN20526  & 57800  & 58715    \\
        1  &  01a & freckles& gold  & STA1600LN-SN25856  & 58719  & 59465     \\
        2  &  01a & fuzzy   & green & STA1600LN-SN20526  & 59466  & 59830    \\
        3  &  01a & wormy   & magenta &STA1600LN-SN31147  & 59830  & -         \\
        \hline 
        4  &  02a & alien & red & STA1600LN-SN19002  & 57800  & 58717     \\
        5  &  02a & alien & red & STA1600LN-SN19002  & 58718  & 59519     \\
        6  &  02a & alien & red & STA1600LN-SN19002  & 59522  & -         \\
        \hline   
        7  &  03a & cruddy & blue & STA1600LN-SN30634  & 59561  & -         \\
        \hline  
        8  &  04a & freckles & gold & STA1600LN-SN25856  & 59605  & -         \\
        \enddata
        \caption{Detector configurations. \rev{The Site corresponds to telescopes as follows: Mauna Loa (MLO) = 01a, Haleakala (HKO) = 02a, South Africa (STH) = 03a, Chile El Sauce (CHL) = 04a. 
        \revtwo{
        The chip ID represents the chip number that we use in this paper to define the combinations and date ranges. The CCD nickname and cryostat color are for ease of remembering the hardware and the formal serial numbers of the devices are listed to confirm the chip provenance.  Note that chips 4,5,6 are all the same configuration, on the same telescope, and allow us to examine the stability of the ATLAS detectors over time. Additionally, chip 8 is physically the same as chip 1 and was moved from MLO to CHL.}}}
        \label{tab:detector_summary}
\end{deluxetable*}

\section{Data preparation}\label{sec:data_prep}
    \begin{figure}
        \centering
        \includegraphics[width=\linewidth]{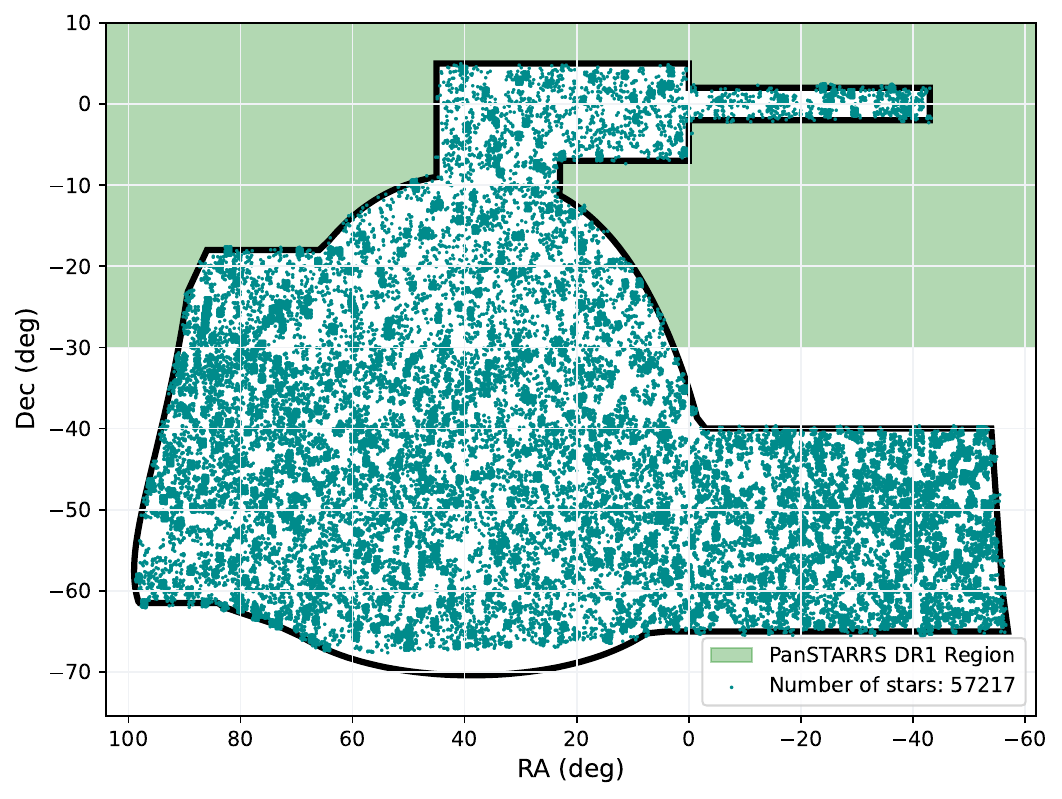}
        \caption{Each individual star field by RA and Dec. The DES footprint is over plotted here along with the Pan-STARRS region. Note that the southern telescopes take over slightly below the PS1 region at Dec of -50$^\circ$. Stars were chosen in 1 square degree chunks randomly distributed throughout the DES footprint. There are 500 chunks each containing roughly 200 stars in our `color-blind' sample, and about 50 stars over 500 chunks in our `blue' sample. In total there are roughly 125,000 stars with full ATLAS history light curves collected, although this number is reduced after cuts described in Sec.~\ref{sec:data_prep}.} 
        \label{fig:skymap_stars}
    \end{figure}

    \subsection{Tertiary Star Samples} \label{sec:tertiary_star_samples}

    For this paper, we build three distinct tertiary star catalogs. First, we construct a baseline sample of stars that are common to both Refcat2 and the DES Y6 catalog \citep{bechtol_2025_DesGold}, \rev{uniformly-distributed in color by resampling the intrinsic color distribution. This re-sampling is important because we measure the slope of our observed - synthetic data residual as a function of color for these stars. Having an even distribution of stars across the entire color range is important to avoid bias or not account for slope at a certain color. This sample is referred to hereafter as the 'color-blind' sample. The color-blind sample has few stars with g - i color $<$ 0.2, with most stars in the blue (g - i color $>$ 0.2)}. Third, \rev{a uniformly-sampled catalog is assembled for only blue stellar colors (DES g - i color $\le$ 0.2) from the common Refcat2 and DES Y6 stars.} This sample is referred to hereafter as the 'blue' sample. We use this blue sample in our calibration because the SNe~Ia primarily exist in this color range  and it enables us to create a uniform in color, total star catalog, for calibration \citep[following][]{Brout19}. Third, a baseline sample of randomly-distributed stars from DES Y6 that are {\textit{not}} found in the Refcat2 catalog. These stars are functionally similar to SNe~Ia, an object whose color and brightness is not used in any part of ATLAS calibration (including initial Refcat2 zeropoint calibration). This sample is referred to hereafter as the 'non-Refcat2' catalog. For all samples we apply cuts as recommended by the ATLAS team in \cite{Tonry_2018_refcat}. We also apply cuts on \revtwo{observations} with excessively large errors ($\sigma_F \ge2000\mathrm{ }\mu\mathrm{Jy}$) or $\chi^2$ above 5, and retain only stars with DES $r$-band magnitudes in the range $17\le r_{\mathrm{DES}} \le19$ mag. Fig.~\ref{fig:skymap_stars} shows the locations of the stellar samples used. This figure only includes stars that pass cuts and are used in calibration ($\sim50,000$). 

    \revtwo{Another aspect considered in the creation of our tertiary star samples is how ATLAS photometry is represented within Refcat2. Because ATLAS photometry is calibrated to the Refcat2 catalog, the surveys contributing to a given Refcat2 magnitude are important. We examine this effect in detail in Sec.~\ref{sec:reference_catalog_validation}. The primary result of this analysis is that stars with Refcat2 magnitudes derived solely from Gaia measurements exhibit significant systematic offsets relative to PS1. To avoid introducing this bias into the ATLAS calibration, we remove all calibration stars that are observed only by the Gaia survey in Refcat2.}

    \subsection{ATLAS Forced Photometry}\label{sec:atlas_fp}

     \begin{figure}
        \centering
        \includegraphics[width=\linewidth]{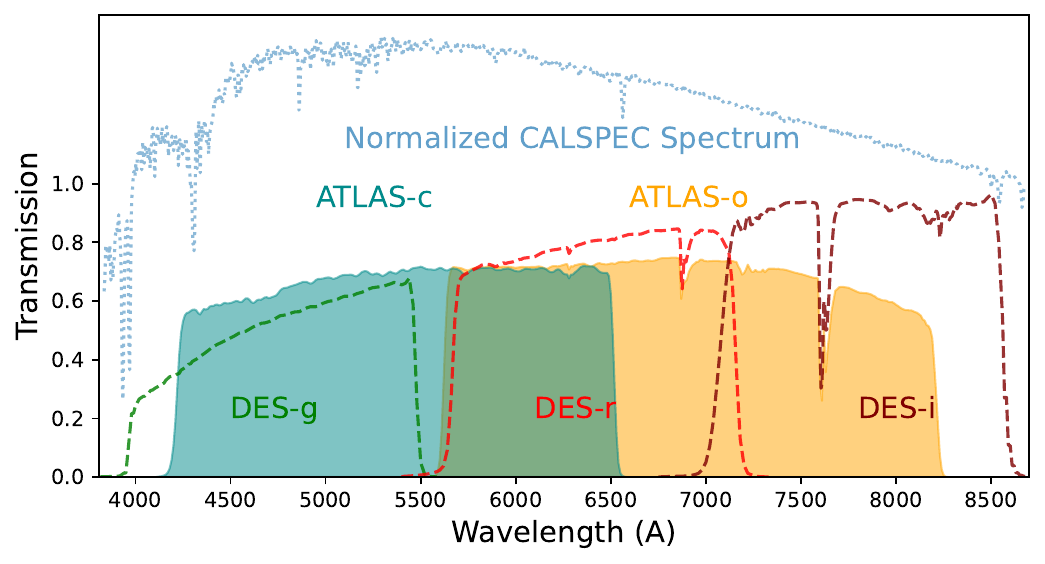}
        \caption{Transmission vs wavelength for ATLAS orange and cyan bands. The DES \textit{g, r, i} bands used for cross calibration are shown for reference. Transmission throughput data comes from SVO2. Also overplotted is HST CALSPEC synthetic star \rev{\textit{hd009051}} used in our calibration. The CALSPEC star's flux density is scaled up arbitrarily,  to be visible on the same scale as the filter functions.} 
        \label{fig:transmision_function}
    \end{figure}

    \begin{figure*}
        
        \centering
        \includegraphics[width=\linewidth]{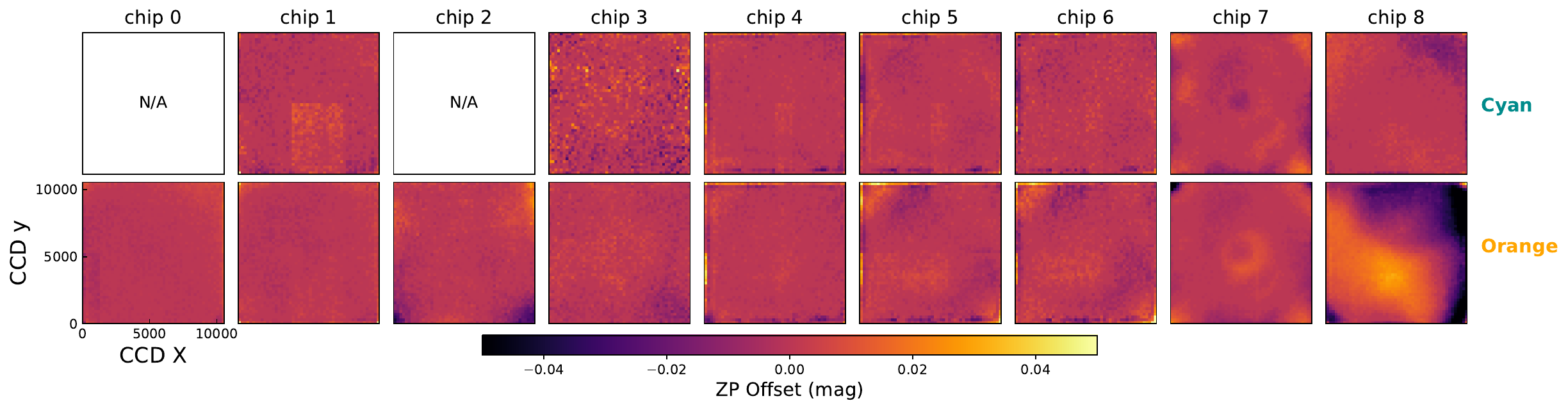}
        \caption{The magnitude residual \rev{within each chip (zero median)}. Median binning applied (50 pixel bins). The heat maps have median residuals of each chip subtracted out. This facilitates characterization of \revtwo{coarse x-y positional variation measurements at the 10s of pixel level}. \rev{See Sec.~\ref{sec:intrachip} for details on heatmap construction.} Note the dramatic variations for chip 8o. \rev{No inter--chip or wave shift correction is applied.}} 
        \label{fig:intrachip_allchip}
    \end{figure*} 
    
    We take our three catalogs and obtain photometric measurements in ATLAS observations by performing the standard forced photometry routine (\texttt{tphot}). \texttt{tphot} is a custom point-spread-function (PSF) fitting routine: it runs on the difference images of ATLAS forced photometry to produce flux measurements. \revtwo{In order to reproduce the same measurement process for the SNe~Ia in the TITAN sample, we measured the photometric fluxes of standard stars in the same way. We used \texttt{tphot} in forced mode and forced PSF fitting at their known positions. This used the same software routines as available on the publicly available ATLAS forced photometry server \citep{atlasserver, Smith20}\footnote{ https://fallingstar-data.com/forcedphot/}}. There is no proper-motion involved in these requests. \rev{Instead, we apply an outlier rejection system and calculate stellar medians instead of means throughout our analysis. With enough sample size we should be able to ignore stars with large proper motion and have enough sample remaining.}

    \subsection{DES Photometry}

    The DES Y6 survey \citep{Rykoff2023} is an incredibly robust ($<$2mmag relative uniformity over the survey region) and well measured survey, covering a large 5000 square degree portion of the sky. Most of the 17 million stars contained within DES Y6 have \textit{i}-band magnitude: $16 <$ \textit{i} $< 21$. The survey uses a modification of Foward Global Calibration Method (FGCM) from \cite{Burke_2018} to remove positional discrepancies across the DECam CCD. 
    
    The absolute calibration is done with the Hubble Space Telescope (HST) CALSPEC standard star \textit{C26202} as specified by \cite{Rykoff2023}. Including systematic uncertainties, DES photometry is calibrated to \textit{C26202} with an accuracy of approximately $1\%$ in flux. 

    \subsection{Synthetic Data}
    
    We generate synthetic ATLAS and DES photometry with NGSL templates \citep{Koleva_2012_NGSL} and CALSPEC standard stars. We take transmission functions for ATLAS from \cite{Tonry_2018_atlas}. We do this by fitting a spectrum of a CALSPEC or NGSL star, to our filter functions wavelength grid. We then integrate this spectrum flux in the photon count space (as opposed to the energy space), and convert this to AB magnitude at the photon pivot wavelength, where AB mag has to be defined in the frequency space. In order to do this at a large scale, we modified the code from \cite{Popovic_2025_Dovekie} to include the ATLAS filter functions. This enables us to produce synthetic stellar photometry for all of our filters at different wavelength shifts quickly. \rev{Our method also allows us to adjust or shift the band pass wavelength if we find discrepancies.}
    
    \subsection{ATLAS CCD - Filter System}
    
    ATLAS's four telescopes, 9 CCDs, and two filters (orange and cyan), result in 18 unique CCD-filter configurations. CCD Chips 0 and 2 never took data in the cyan band, leaving 16 total CCD-filter configurations. The four telescopes that comprise ATLAS began operating about a decade ago with the first northern telescope starting operation in June 2015 (HKO), the second in February 2017 (LMO), the two southern telescopes began operation in 2021. \rev{We chose to start the TITAN data sampling, and the calibration data, in early 2017 (MJD=57800), at a time when the northern ATLAS units had settled down to a stable operating mode and hardware configuration.} Fig.~\ref{fig:transmision_function} shows the flux density of each ATLAS and DES filter as a function of wavelength, with a reference CALSPEC stellar spectrum. We observe that ATLAS's coverage approximately lines up with DES \textit{g,r,i} bands. 

    Another key note is that the quantum efficiency (QE) is not uniform \revtwo{between the CCDs used in the two northern telescopes (Tab.~\ref{tab:detector_summary} shows the changes in CCDs)}. It is largely uniform until $\lambda = 6500\mathrm{\AA}$, where there is deviation over the rest of the wavelength we use. We do not attribute substantial effects in our calibration with QE. See Fig.~3 in \cite{Tonry_2018_atlas} for additional details about QE in ATLAS.

    \subsection{ Refcat2 Catalog Validation} \label{sec:reference_catalog_validation}

    For each image we collect from ATLAS there is a zeropoint calculated using stars from the Refcat2 catalog. This catalog is a combination of many different surveys to facilitate all-sky coverage for ATLAS. The primary surveys involved here are PS1 \citep{Flewelling_2020_PS1} and Gaia \citep{Gaia_2018_gaia}, \rev{with GAIA, APASS DR9 \citep{Henden_2016_APASSDR9} and Skymapper DR1 \citep{Wolf_2018_Skymapper} in the south} \citep{Tonry_2018_refcat}.

    First, we aim to validate that no single survey from Refcat2 is providing chromatic or skewed data, thus biasing ATLAS photometry. The Refcat2 catalog combines every survey that measures a star's magnitude and averages them together. There is no clean way to determine \rev{a single survey's contribution to the Refcat2 magnitude value when multiple surveys observe a star}. Thus, to conduct this validation, we examine Refcat2 stars that 1) only have contribution from one survey, or 2) are specifically missing contribution from one survey. \rev{This allows us to isolate effects that might occur from each survey individually. Our primary finding is that stars only measured by Gaia are skewed substantially off the main PS1 survey (above declination of -40).} To avoid this potential bias in ATLAS photometry, we filter out all calibration stars that only have observations by the Gaia survey. 

    Other than the discrepancy with Gaia, the rest of the surveys match well with the trend of PS1, including the other surveys \rev{in the south where PS1 data does not exist}. This is facilitated by the overlap from the highly-uniform DES Y6 catalog facilitating comparison. \rev{We remove stars that only have Gaia measurements in Refcat2 from our calibration.} 

\section{Analysis} \label{sec:analysis}

    \begin{figure}
    \centering
    \includegraphics[width=\linewidth]{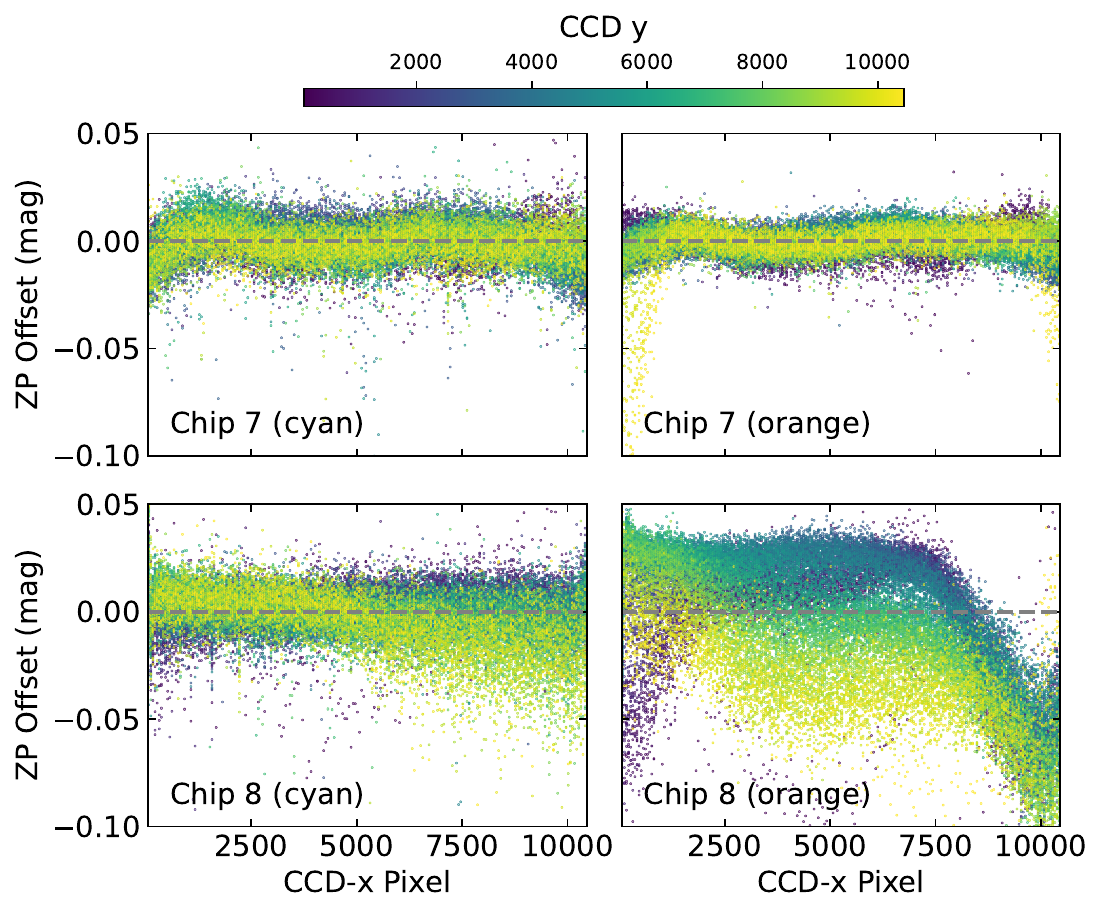}
    \caption{Collapsed 1D views of two chips \rev{(7,8)} from Fig.~\ref{fig:intrachip_allchip} in orange and cyan bands. The pixel offsets in magnitude are shown on the y-axis and the x-axis is the x pixels of corresponding Fig.~\ref{fig:intrachip_allchip}. This shows the significant non uniformity of chip 8o as a function of x pixel. Note how in cyan, while variations are larger for chip 8 compared to chip 7, there are no significant deviations from uniformity. \rev{Chip 7 is representative of a more typical chip used in this analysis and chip 8 is highlighted as an area for future improvement and ongoing work.}}
    \label{fig:binned_pixel_map}
    \end{figure}
    
We break our calibration down into two primary components. First, we have the intra--chip calibration, where we have examined CCDs of each telescope repeatedly to determine \revtwo{trends within the CCD at the binned (10s of pixels) pixel level} that can be corrected and facilitate better nightly precision. \revtwo{This provides a coarse x-y positional dependence measurement.} Second, is inter--chip calibration. We examine trends across all filter-chip combinations to produce a median $\Delta$ zeropoint \revtwo{($\Delta$ZP)} offset for each individual chip-filter. This portion also includes shifting any filters in wavelength to correct for chromatic effects. We conduct this filter shift in a phenomenological manor, focusing primarily on optimal calibration for SN~Ia cosmology.

    \subsection{Intra--Chip Variation} \label{sec:intrachip}

    \begin{figure}
    \centering
    \includegraphics[width=\linewidth]{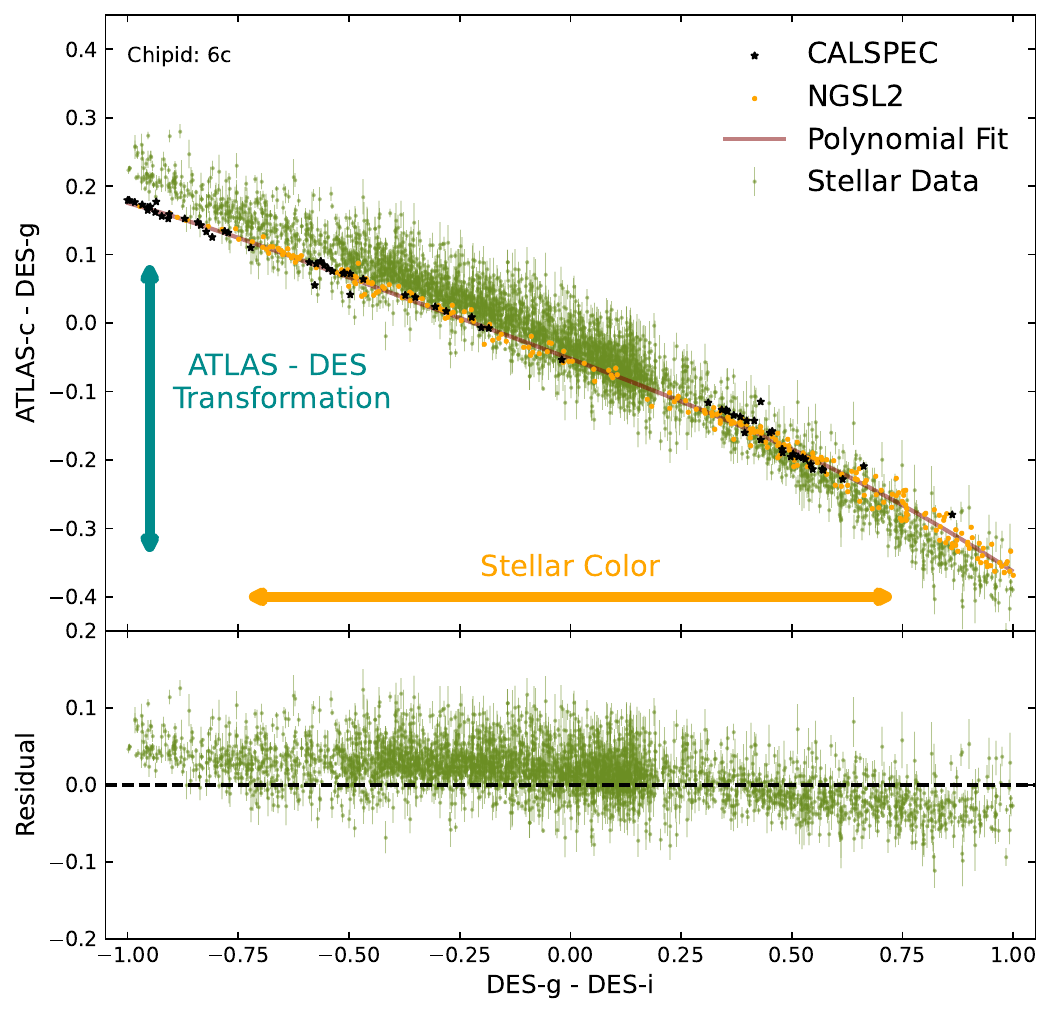}
    \caption{ATLAS - DES transformation (ATLAS-c - DES-g) versus DES color \rev{(g - i)} for ATLAS chip 6c. Real stellar data from ATLAS server and DES Y6 is shown in green. Both NGSL and HST CALSPEC synthetic stellar photometry (orange, black) are overplotted. A 5th order polynomial is fit to the synthetic \rev{NGSL} data (brown) is shown. Lower plot is the real data residual to the polynomial fit. The lower residual plot demonstrates a residual chromatic slope that must be accounted for. The net vertical shift of the green points relative to the trend line shows the inter--chip offset. The actual calculation of these offsets are substantially more complex than what is shown here and the likelihood and fitting process are described in detail in App.~\ref{appendix:multi-color_joint_likelihood}. \rev{This inter--chip correction is applied \textit{after} accounting for intra--chip variation.}} 
    \label{fig:offset_plot_example}
    \end{figure}
    
    The untargeted, all-sky survey pattern by ATLAS creates a dither pattern around each star's coordinate. This pattern provides an insight into the sensitivity function's possible variations within each CCD, as tertiary standard stars are measured at many different CCD coordinates \rev{and across the focal plane}.

    ATLAS CCDs have 10560X10560 pixels \rev{(STS - 1600 model)}, and each image is read out in 1x1-binning by default. \rev{Median seeing is $3.7'' - 5.6''$, that span 2 - 3 pixels at full width half max (FWHM), with each pixel containing $1.86''$.} The \texttt{tphot} forced photometry routine reports the CCD coordinates ($x$,$y$) that correspond to the requested sky coordinates for the forced photometry. We use this information to construct the coordinate-dependent zeropoint offset map within each chip.

    The procedure is the following: for any given star, we have multiple observations across different x,y coordinates in multiple CCDs. Then for any star with data in a given CCD, we take the median of all magnitude values, and subtract that from each individual \revtwo{observation} magnitude value. This creates a coordinate dependent offset of one star mapped across all chips. \revtwo{We then repeat for every star producing a heatmap of the coordinate based offset within a CCD.} Fig.~\ref{fig:intrachip_allchip} shows the results of this process. Because we are looking for a coordinate dependence, we subtract out the median offset from each chip, to make net offset 0 if there is no coordinate dependence. \rev{As shown in \cite{Bernstein_2017},} we are ignoring edge effects on all the chips as those are notoriously unreliable across CCDs, thus they are cut out, at the 50 pixel scale, before correcting. 
    
    Fig.~\ref{fig:intrachip_allchip} shows chips 0-7 have no particularly concerning patterns, i.e. \revtwo{variation at the 10 pixel level}. We can see some distinct patterns on the \rev{1 mmag} level. Since these are different filters and thus, data in one filter is independent of data in another, this is a strong validation that these patterns (and thus those more significant like chip 8o) are physical results, and not a product of our data processing. 
    
    \begin{figure}
        \centering
        \includegraphics[width=\linewidth]{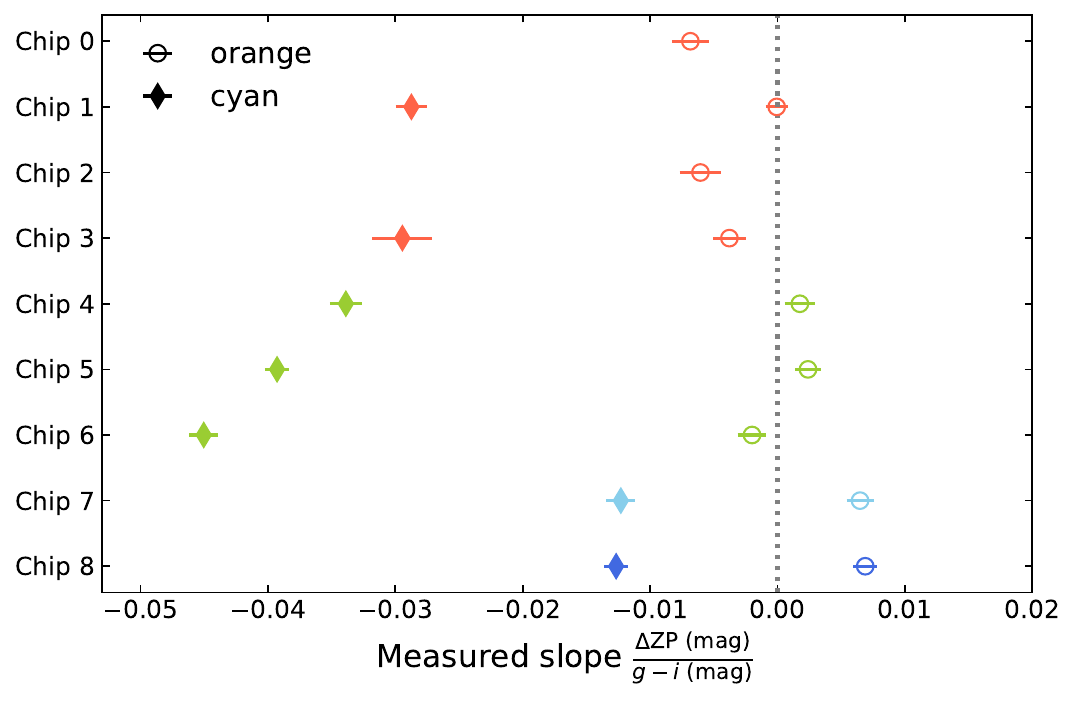}
        \caption{The measured slopes ($\Delta$ZP) of the residuals to polynomial fits (illustrated for chip 6 at the bottom of Fig.~\ref{fig:offset_plot_example}) for each chip (y-axis). An example of the slope measured can be seen visually in the bottom panel of Fig.~\ref{fig:offset_plot_example}. Points are colored by which telescope each chip corresponds to \rev{in the order presented in Tab~.\ref{tab:detector_summary}}. The larger the slope value, the larger the wave shift we apply, thus, this plot, indicates the significance of the shift. \rev{Actual shift values presented in Tab.~\ref{tab:offset_comparison}}.}
        \label{fig:slope_whiskers}
    \end{figure}

    Chip 8o has a significant vignetting pattern with brighter magnitude residuals toward the right side and slightly at the top of the chip. Fig.~\ref{fig:binned_pixel_map} also shows a scatter plot of the x and y axes of chip 7 on the top plots, and the same plot for chip 8 on the bottom plots. Clearly visible here is the trend in the x axis of the chip toward brighter observations on the right. A significant observation from Fig.~\ref{fig:binned_pixel_map} is that this vignetting pattern, producing brighter observations in chip 8o, only exists in the orange band.

    We account for this vignetting pattern in our correction model. We create the correction model by binning the pixels of each chip into 50 pix bins. We use our calculated 'optimal smoothing radius' of 540 pixels (App.~\ref{appendix: optimal_smoothing_radius}), to convolve our 2D arrays using python's \texttt{Gaussian2DKernel}. This convolution then gets remapped to the entire 10,560 by 10,560 pixel space to produce a complete correction map for one chip. Our model applies unique corrections to each chip and each filter separately (16 total correction maps). These maps are then \rev{combined sequentially} and applied to our calibration.  
    
    \subsection{Inter-Chip Variation} \label{sec:inter-chip}
    \begin{figure}
        \centering
        \includegraphics[width=\linewidth]{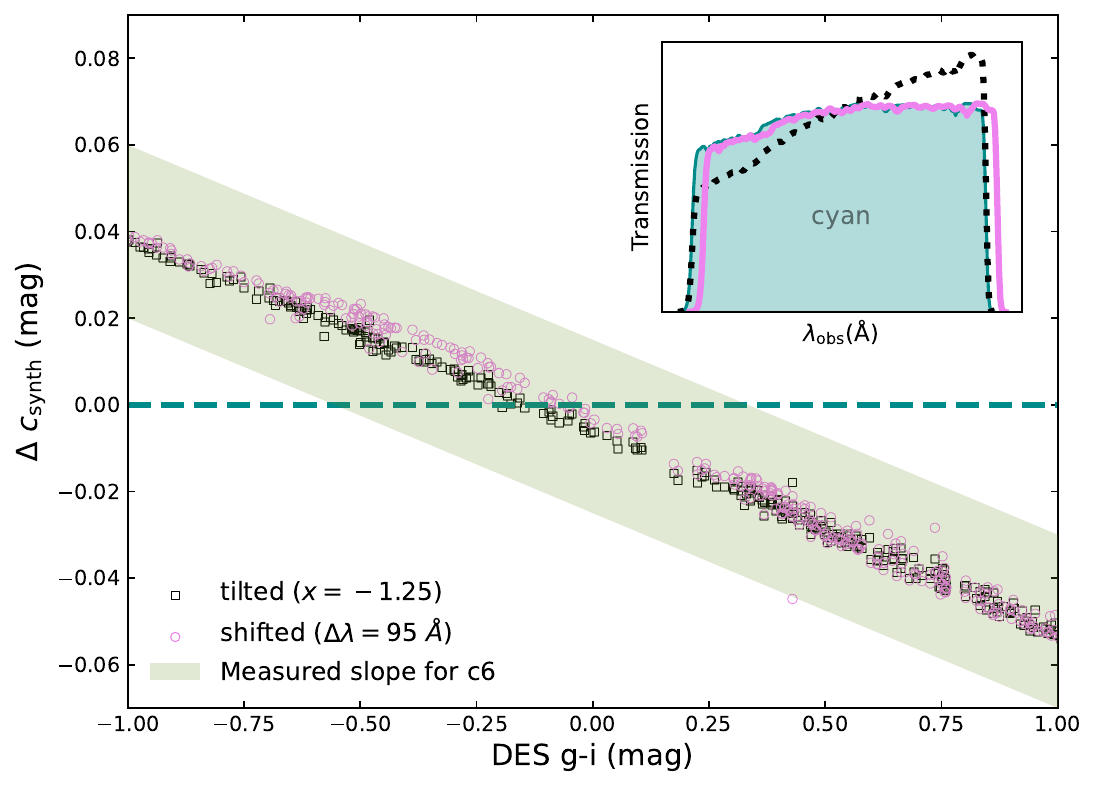}
        \caption{The y-axis is the synthetic magnitude of cyan band with no shift minus the magnitude with a wavelength shift. The  x-axis is DES g - i color. Both 95 \AA. coherent shift (pink) and a tilted throughput with $X=-1.25$ (black; see Eq.8 in \citealt{Popovic_2025_Dovekie} for definition) are shown for comparison. The green shaded region shows the observed tilt for chip 6c, the most extreme case shown in Fig.~\ref{fig:offset_plot_example}. The subplot in the top right corner shows the original, shifted, and tilted transmission functions.}
        \label{fig:filter_shift}
    \end{figure}

    The inter--chip offset is described as the vertical shift between the observed ATLAS - DES transformation function and the synthetic transformation function in Fig.~\ref{fig:offset_plot_example} for each chip. The residual plot on the bottom of Fig.~\ref{fig:offset_plot_example} shows the value of this vertical shift as a function of DES color. The residual is calculated as the y axis difference between the real data at that color and the value of a polynomial fit to the synthetic NGSL2 photometry. The synthetic polynomial is a 5th order approximation of the synthetic data using python's \texttt{Polynomial.fit}. 

\begin{figure}
    \centering
    \includegraphics[width=\linewidth]{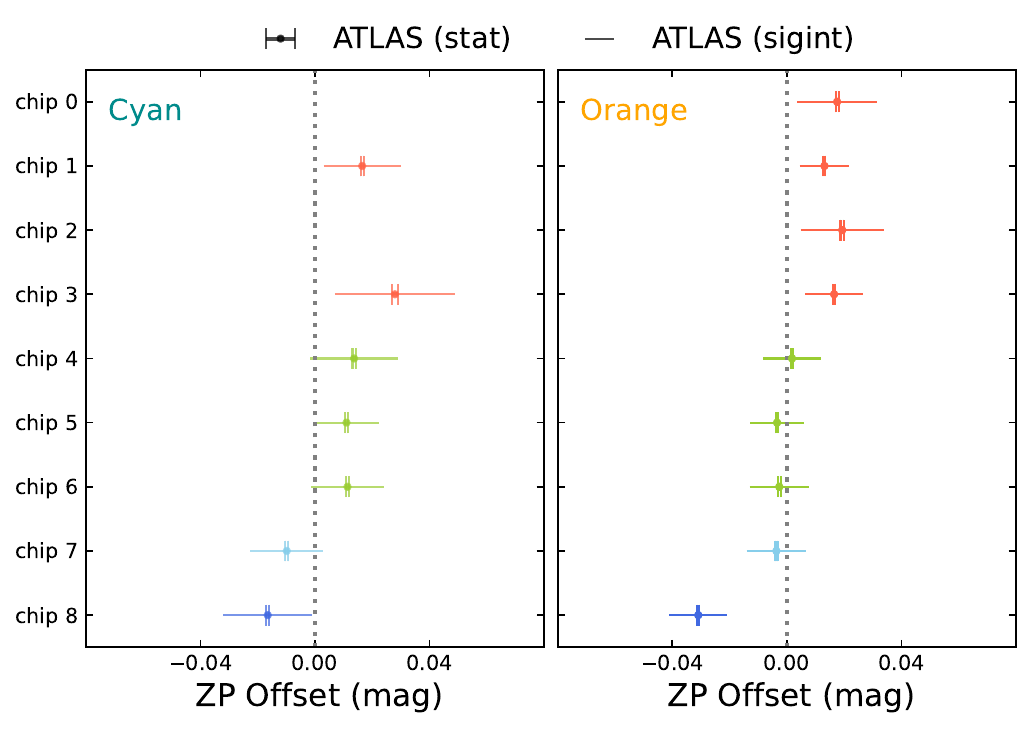}
    \caption{\rev{This figure summarizes the inter--chip zeropoint corrections applied for each chip.} These offsets are the zeropoint shift of the real data from the synthetic data in ATLAS-DES transformation vs color, as demonstrated in Fig.~\ref{fig:offset_plot_example} and calculated following Sec.~\ref{appendix:multi-color_joint_likelihood}. Two error bars are displayed: 1) the smaller errors represent the statistical (`stat') uncertainty resulting from the ATLAS data, and 2) the larger errors represent the dispersion of the data (`sigint').}
    \label{fig:offset_chip_comparison}
    \end{figure}

    We expect that this residual is a flat line centered away from 0. The amount this line is offset from 0 would be the zeropoint offset of this chip-filter combination (there is a collapsed likelihood function used here to generate this, but it is still the result of this residual). This is the zeropoint offset because the synthetic data uses CALSPEC stars, which have the absolute flux of our filter function. \revtwo{Fig.~\ref{fig:offset_chip_comparison} shows the results of this zeropoint offset, these are the values that are applied to each respective chip during the inter--chip correction.}

    Notably, there is a $g-i$--color dependent trend in the residuals \rev{(the bottom plot of Fig.~\ref{fig:offset_plot_example} shows the most egregious case, most chips are substantially better)}. Considering that this is the residual of observed photometry from the synthetic photometry, the presence of a slope implies that our filter transmission functions used for the synthetic photometry differ from each telescope-detector-filter combinations' actual throughput. \revtwo{Fig.~\ref{fig:slope_whiskers} shows this chromatic effect slope across all chips. Note the systematic chromatic effect in cyan band vs the minimal effect in orange. Also note that because chips 4,5,6 are all the same camera setup over time, we can use this to observe that the chromatism worsens with time even within a singlechip chip.}
    This is not an unusual observation: previous cosmology-grade calibrations of SNe~Ia catalogs, such as \cite{Brout22} and \cite{Popovic_2025_Dovekie} have identified chromatic slopes using a similar method. Unless a careful, laboratory-level re-measurement of the system throughput can be performed, these slopes are typically corrected by applying modifications to each filter's transmission function.
    Fig.~\ref{fig:filter_shift} demonstrates the color-dependent (chromatic) effect of such modifications: two distinct methods (wavelength--shift and filter--tilt; see \citealt{Popovic_2025_Dovekie} for review) produce a nearly identical color--dependent change in the predicted magnitudes. For consistency with the literature and simplicity, we choose to employ the wavelength-shift method. A correct choice of wavelength shift can match the measured slope in the tertiary stars, effectively mitigating the chromatic effect during the light-curve fitting of the cosmological SNe~Ia samples.

    The measured slopes, and therefore implied filter shifts are most pronounced for cyan where we find each chip should be shifted by 50 - 100 \AA\xspace in the same direction. Although this initially sounds substantial, given how broad the ATLAS filter bands are this is comparable to shifting a DES or PS1 band by 25 - 50 \AA, which has been shown to be necessary in some cases (e.g., PS1-$g$; \citealt{Scolnic_2015}). 
    While the exact cause of the observed chromatic effect is unknown (e.g., change in quantum efficiency, filter degrading, calibration issue)
    our phenomenological approach is efficient at removing the observed chromatic effect, and is backed up in the literature as a viable solution for cosmology. \rev{Furthermore, in Section \ref{sec:validation_with_calspec}, we will demonstrate the validity of these cyan shifts on the independent CALSPEC spectrophotometry.}

\begin{deluxetable}{ccccc}
    \vspace{0.3cm}
    \tablecaption{Orange and cyan band zeropoint offset corrections and wavelength shifts for TITAN calibration. Zeropoint offset corrections are in units of magnitude while wavelength shift is in units of \AA. Note that both $\Delta m$ and $\Delta \lambda$ need to be used as a set, as $\Delta m$ is calculated as an offset from a synthetic photometry using the corresponding $\Delta \lambda$ for each chip.}
    \label{tab:offset_comparison}
    \tablehead{
        \colhead{Chip} & 
        \colhead{$\Delta m_\mathrm{cyan}$} & 
        \colhead{$\Delta \lambda_\mathrm{cyan}$} & 
        \colhead{$\Delta m_\mathrm{orange}$} & 
        \colhead{$\Delta \lambda_\mathrm{orange}$}   \\
        &
        \colhead{(mag)} & 
        \colhead{(\AA)} & 
        \colhead{(mag)} & 
        \colhead{(\AA)} 
    }
    \tabcolsep=2.5pt
    \startdata
    0 & \nodata & \nodata & $+0.176 \pm 0.0005$ & $+22 \pm 2.7$ \\
    1 & $+0.017 \pm 0.0005$ & $+56 \pm 1.4$ & $+0.013 \pm 0.0003$ & $+5 \pm 1.7$ \\
    2 & \nodata & \nodata & $+0.019 \pm 0.0006$ & $+25 \pm 3.1$ \\
    3 & $+0.028 \pm 0.0011$ & $+57 \pm 2.4$ & $+0.017 \pm 0.0005$ & $+27 \pm 2.8$ \\
    4 & $+0.014 \pm 0.0006$ & $+67 \pm 1.4$ & $+0.002 \pm 0.0004$ & $-6 \pm 2.1$ \\
    5 & $+0.011 \pm 0.0004$ & $+78 \pm 1.1$ & $-0.003 \pm 0.0004$ & $-6 \pm 1.9$ \\
    6 & $+0.011 \pm 0.0005$ & $+87 \pm 1.6$ & $-0.003 \pm 0.0004$ & $+10 \pm 2.4$ \\
    7 & $-0.010 \pm 0.0005$ & $+28 \pm 1.1$ & $-0.004 \pm 0.0004$ & $-15 \pm 1.8$ \\
    8 & $-0.017 \pm 0.0005$ & $+28 \pm 1.0$ & $-0.031 \pm 0.0004$ & $-21 \pm 1.8$ \\
    \enddata

    \vspace{-0.8cm}
\end{deluxetable}

\section{\rev{TITAN Calibration Validations and Tests}} \label{sec:validation_tests}

    \rev{We validate our calibration in three ways. Against an independent tertiary star catalog ('non-Refcat2' catalog from Sec.~\ref{sec:tertiary_star_samples}), using HST CALSPEC primary calibrators and DA white dwarfs, and analysis of coordinate dependence of the tertiary star residuals before and after correction.}

\begin{figure}
    \centering
    \includegraphics[width=\linewidth]{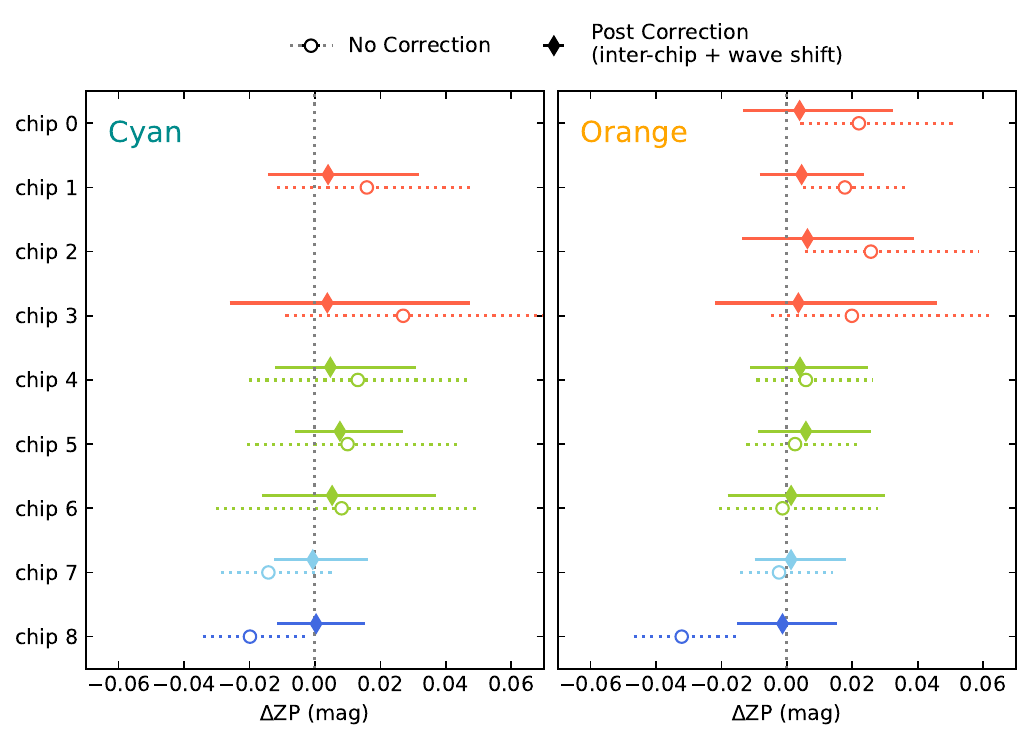}
    \caption{\rev{$\Delta$zeropoint offset in magnitude between observed and synthetic photometry (same definition as Fig.~\ref{fig:offset_chip_comparison}) for a validation sample of stars which are independent of Refcat2, before and after our calibration correction.} 
    The $\Delta$ZP value is the offset between chips and is corrected by the inter-chip correction. The error on the values is the result of a projection of the residuals from Fig.~\ref{fig:offset_plot_example} into the $\Delta$ZP offset space, representing the scatter in residual. This error is dominated by the chromatic slope displayed in Fig.~\ref{fig:slope_whiskers}. 
    }
    \label{fig:offset_validation}
\end{figure}

\subsection{Validation with Independent Tertiary Catalog} \label{sec:validation_independent_tertiary_catalog}
     We first validate using tertiary stars that are not contained in the Refcat2 catalog and therefore are not used in our calibration solution. These stars are identified in the DES Y6 catalog for which ATLAS forced photometry is obtained as outlined in Sec.\ref{sec:atlas_fp} (this is what is referred to as the 'non-Refcat2' catalog in Sec.~\ref{sec:data_prep}). This provides an independent photometric dataset for validation. From the perspective of ATLAS, these 'non-Refcat2' stars behave functionally the same as SNe~Ia: a point source object that is not included in zeropoint calibration of each image. Fig.~\ref{fig:offset_validation} displays the effect our calibration has on these validation stars. Additionally, the error bars on these points represent the percentile range (16th percentile to 84th percentile) of the median value offset of a chip.

    Fig.~\ref{fig:offset_validation} shows the results for the 'non-Refcat2' stars before and after our calibration solution. First we find that the scatter in stars (16th percentile to 84th percentile error bars) is reduced, especially in cyan band, is reduced. Second, after correction, all of zeropoint offsets relative to DES are near zero. Note chips 4c, 5c, 6c, all of whose stellar scatter is reduced substantially with correction, which is the result of accounting for the chromatic slopes.

    \begin{figure}
        \centering
        \includegraphics[width=\linewidth]{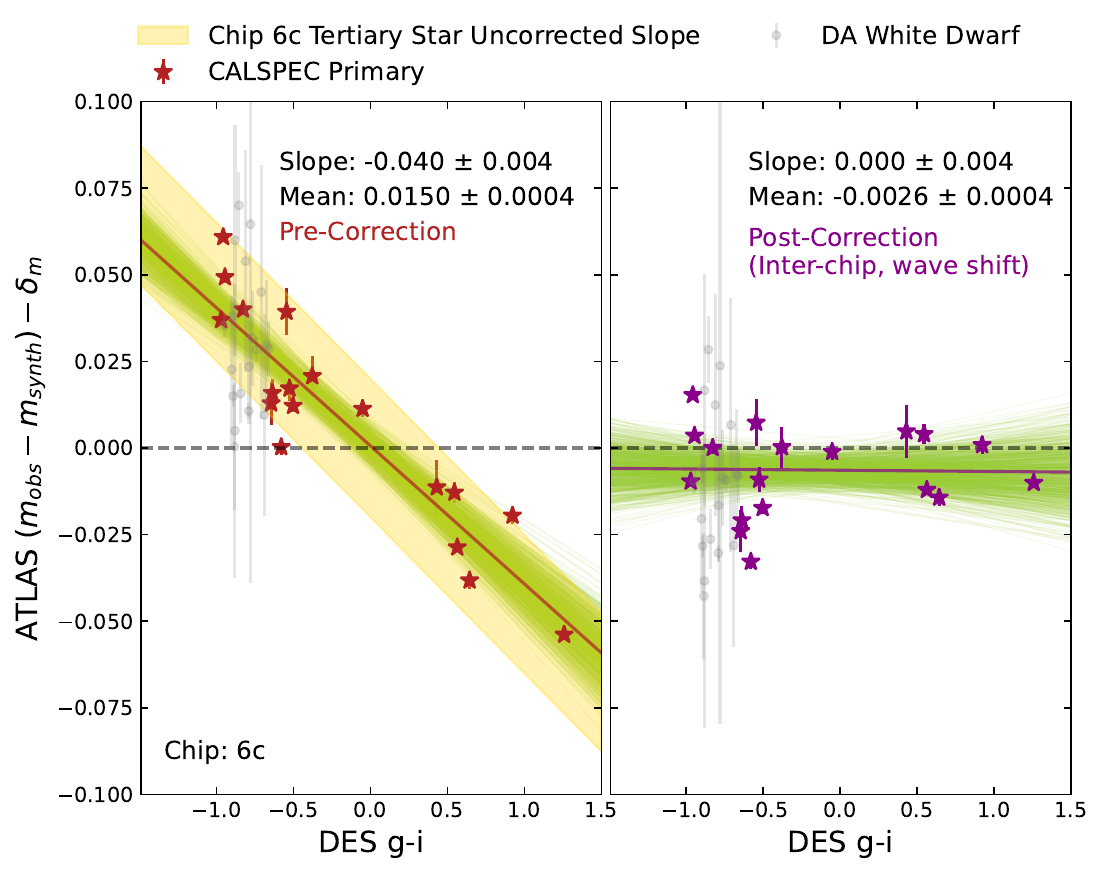}
        \caption{Synthetic minus observed residuals of several CALSPEC stars, including C26202, and DAWD, for ATLAS cyan band (chip 6c) versus DES color in $g - i$. \rev{This is nearly identical to the bottem panel of Fig.~\ref{fig:offset_plot_example} but now demonstrating the impact of our calibration corrections}. The many green lines represent random slope draws from the likelihood fit that account for covariance to show a range of possible fitted slopes and uncertainty. The solid lines are the best-fit slopes. We note that values for all chips are reported in Tab.~\ref{tab:primary_slope_offsets}}.

        \label{fig:primary_calib}
    \end{figure}

    \begin{figure*}
        \centering
        \includegraphics[width=\linewidth]{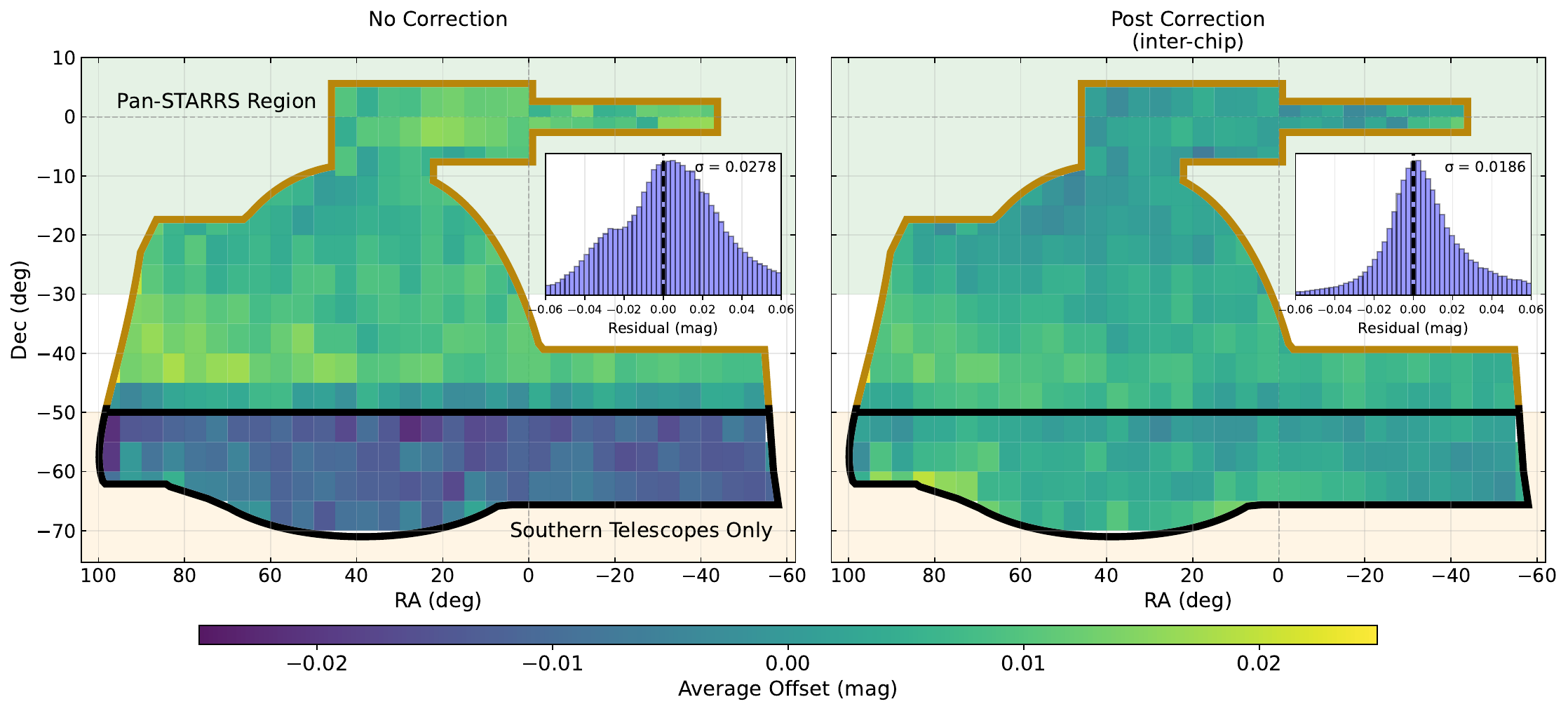}
        \caption{Heatmap of median offset as a function of spatial position before and after inter-chip corrections. After corrections,  the spatial dependence and the standard deviation of the residuals improve. This is most noticeable around -50$\degree$ declination where the northern telescopes cut off. The histogram residuals between ATLAS and DES stars after corrections becomes substantially more gaussian.}
        \label{fig:median_offset_map}
\end{figure*}
    \subsection{Validation with HST CALSPEC \& DA White Dwarf Reference Stars} \label{sec:validation_with_calspec}

    The second validation method we employ is using primary, and secondary stars to reproduce Fig.~\ref{fig:offset_validation}. We use a combined dataset of spectroscopic flux-calibrated standards, HST CALSPEC  \citep{Bohlin14} and DA-type faint White Dwarfs \citep[hereafter DAWD;][]{Boyd_2025_DAWD}, to further validate our results and quantify systematic uncertainties. These spectroscopic standards, observed by \textit{HST/STIS} with an absolute calibration to physical units, provide a direct comparison of synthetic and observed spectra without the need of deriving the synthetic color-color transformation (Eq.~\ref{eq:offset_def}). This independency, along with the broadly accepted use of the CALSPEC stars for photometric calibration, makes them an excellent probe to test the possible systematics in the post-correction photometry of the TITAN dataset.

    In addition to enabling an independent check of our tertiary star--based methods, the use of CALSPEC stars is relevant to the original calibrations of DES and ATLAS.
    DES uses a single primary calibrating star's spectra for its calibration to the absolute AB magnitude system, \textit{HST CALSPEC C26202}. DES claims that, including systematic errors, the absolute flux is known at approximately the $1\%$ level. DES generates these synthetic magnitudes by integrating the official DES passband throughputs with one of standard spectra for C26202 from the HST CalSpec database \citep{Bohlin14}.
    PS1, on the other hand, does not use a single CALSPEC star for its absolute calibration: instead, they rely on Ubercal from \cite{Schlafly_2012_ubercal} for the initial zeropoint calibration, which is then tied to the physical units using multiple CALSPEC standards \citep{Magnier_2020_ps1}.

    In Fig.~\ref{fig:primary_calib}, we present the measured offset between synthetic and ATLAS--observed photometry of chip c6 using CALSPEC and DAWD stars. An additional mmag-level offset $\delta_m = m^{DES}_{C26202} - m^{synth}_{C26202}$ is subtracted to account for the difference between DES photometry of their absolute-scale calibrator, C26202, and our synthetic photometry. This is possibly due to the small, numerical effect from the difference in sub-sampling along the wavelength axis. 
We calculate a slope and offset $\Delta\mathrm{ZP}_i=\bar{\mathcal{A}}\cdot(g-i)_i + \mathcal{B}$ (the purple line in Fig.~\ref{fig:primary_calib}) in our post-correction residual. We use the values of the slope ($\bar{\mathcal{A}}$) and the intercept ($\mathcal{B}$) to quantify systematic uncertainties (Section~\ref{sec:systematics}).

    We see in Fig.~\ref{fig:primary_calib} that, our corrections improves the offset and chromatic effect. \rev{The mean is reduced from 0.015 pre-correction to -0.0026 post-correction.} The chromatic slope is reduced from  -0.0399 pre-correction to -0.0004 post-correction. This improvement in slope and offset is an independent validation of the methodology using tertiary star cross-calibration with DES. We show chip 6c as an excellent example validation of our filter-shift correction. We do not present any orange band data for primary calibrators in this plot, orange band data already has minimal slope and corrections are extremely small \rev{(we find an uncorrected median slope of 0.004 in orange band)}. 
    We will discuss the resulting reduction of systematic uncertainty in Sec.~\ref{sec:systematics}.
    
     \subsection{Coordinate Dependence}
    When calibrating four independent telescopes it is important to verify there is no residual coordinate dependence (due to the different physical locations of the telescopes). There are two main regions we might expect coordinate dependence: above/below -50$^\circ$ declination where the northern telescopes cut off, and above/below -30$^\circ$ declination where PS1 (the primary calibrating instrument of ATLAS Refcat2) cuts off. 
    
    In Fig.~\ref{fig:median_offset_map} we can see, before our DES cross-calibration, there is a coordinate dependent offset at the northern telescope cutoff (the bluer region below -50$^\circ$ dec). We are able to remove this offset with our inter--chip corrections as seen in the right side of the plot. \rev{Also apparent is a slightly less defined discrepancy at dec of $-45^\circ$ where the bore sights of the northern telescope pointing positions are set. Beyond this we see no effect at the boundary of the PS1 region, implying that APASS and Skymapper in the south are sufficient calibration catalogs. Therefore, the right side of Fig.~\ref{fig:median_offset_map} shows that applying our inter--chip  correction creates uniformity across the entire DES footprint, and specifically resolves the issues with the southern  telescope calibration.} In the histograms in Fig.~\ref{fig:median_offset_map}, we find the scatter between ATLAS and DES photometry after transformation reduces substantially after our correction (from 0.028mag to 0.019mag), \rev{and the histogram of all ATLAS-DES tertiary comparisons becomes more Gaussian.}

\section{Distances and Hubble Diagram Residuals} \label{sec:dinstance_comparison}

\rev{We apply the calibration defined in this work to the light curves of TITAN DR1 gold \citep{Murakami_TITAN_DR1}, hereafter DR1. In DR1 light curves are fit with the} SALT3-DESY5 model using the SNANA package \citep{snana}. \rev{Our intra--chip, inter--chip, and wave shift corrections are applied specifically during SALT3 fitting, see Fig.~\ref{fig:flowchart} for details on the calibration application.} This fit yields stretch ($x_1$), color ($c$), and B-band model magnitude ($m_B$), and the time of maximum (PKMJD) for each SN. Note that there is a cut on color error at $\sigma_c<0.1$, for details and discussion on the light-curve fits and model residuals, see \cite{Murakami_TITAN_DR1}.

We compare the fitted light-curve parameters ($x_1,\ c$) against the fits to the light curves of the same, cross-matched SN observed by external surveys (DEBASS, YSE, and ZTF).
Additionally, we apply a simple standardization using the SALT2mu routine \citep{Marriner_2011_SALT2mu}, which finds an optimal set of coefficients for stretch-luminosity relation ($\alpha$) and color-luminosity relation ($\beta$), as well as a few additional nuisance parameters. Using an arbitrary absolute magnitude $M_B$, this yields a standardized, distance modulus (solely for the purpose of one-on-one comparison):
\begin{equation}
    \mu_\mathrm{test} = m_B + \alpha\cdot x_1 - \beta \cdot c - M_B\ .
\end{equation}
For the purpose of direct, one-to-one comparison of cross-matched SNe across surveys, no bias correction is needed \rev{and nor is the typical `mass step'}, and we use the same set of ($\alpha$, $\beta$) across all surveys.

    \begin{figure}
        \centering
        \includegraphics[width=\linewidth]{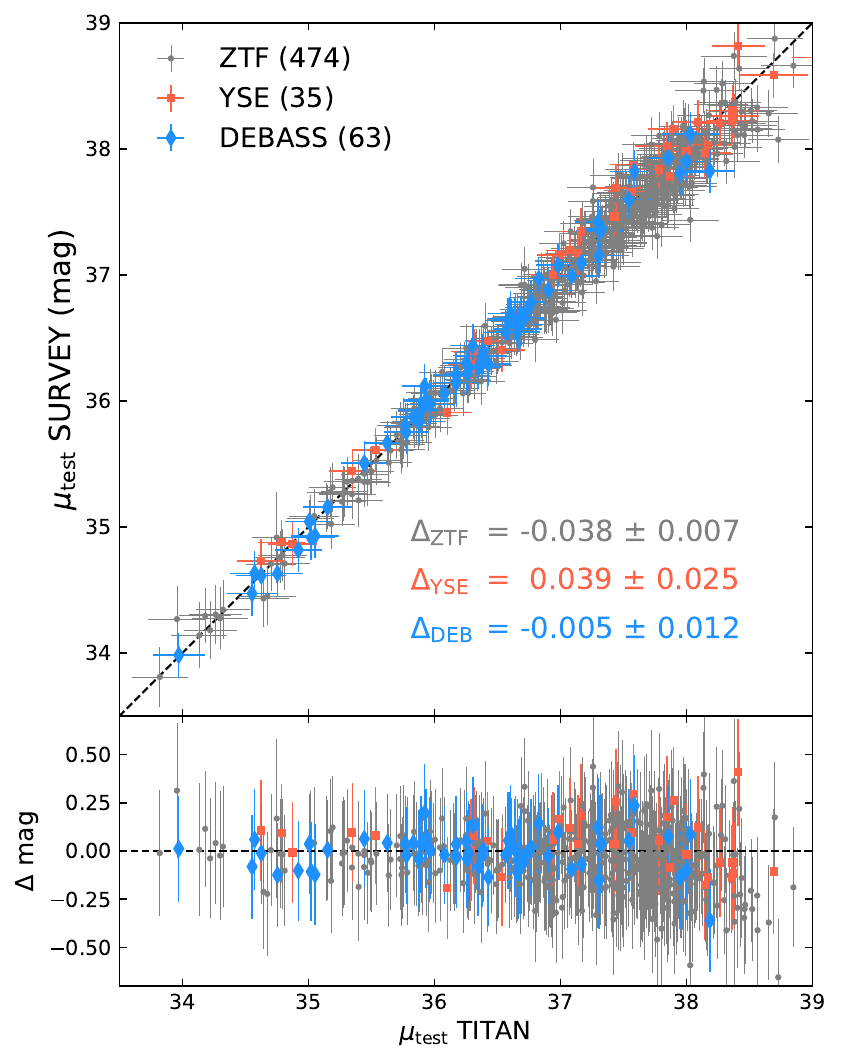}
        \caption{\rev{Fiducial distance moduli $\mu_\mathrm{test}$ for SNe included in both TITAN DR1 and other modern low-$z$ surveys. TITAN light curves are presented in the companion paper \cite{Murakami_TITAN_DR1}. Compared against ZTF DR2, DEBASS DR1, and YSE DR1. We use a conservative color error cut in TITAN SALT3 color, $\sigma_c \le 0.1$, to minimize the known error-dependent bias (see \citealt{Murakami_TITAN_DR1} for discussion). The measured offsets are consistent with zero for YSE and DEBASS. The significant offset between ZTF and TITAN is consistent with the known offset for ZTF reported in \cite{Newman_2025_redgiant}.}
        }
        \label{fig:titan_survey_mu}
    \end{figure}
    
Fig.~\ref{fig:titan_survey_mu} shows a comparison between TITAN and the three other low-$z$ surveys in stretch, color, and distance. We describe the data from additional surveys used in this comparison (ZTF, DEBASS, and YSE) and discuss the implications in the following sections.

\subsection{ZTF} \label{sec:ZTF}

The second data release of ZTF (Zwicky Transient Facility) SNe~Ia sample contains 2667 spectroscopically confirmed Type Ia SN with matching redshifts in the low-$z$ region (z $<$ 0.3) that pass initial cosmology cuts \cite{Rigault_25_ZTFRelease}. \revtwo{We find 474 cross matches with TITAN.} This is one of the largest spectroscopically-confirmed low-$z$ supernova datasets to date. We compare to ZTF as the only other low-$z$ sample with SNe counts on the same order of magnitude as TITAN. Note that ZTF claims to have not completed their calibration for cosmology \rev{due to an observed `pocket effect' of flux-dependent point spread function biases. The ZTF group also noted in \cite{lacroix2025ztfsneiadr2} that an offset in the DR2 magnitude values on the order of $0.09$ mag is needed to correct the overestimated flux in ZTF DR2 \citep{Rigault_25_ZTFRelease}. A separate group, \cite{Newman_2025_redgiant}, finds ZTF to be too bright by 0.024mag in comparison with 28 SNe~Ia in common from Las Cumbres Observatory (LCO) $gri$ photometry.} 

\subsection{DEBASS}

DEBASS (Dark Energy Bedrock All Sky Supernova program) has collected the largest ($>500$ SNe~Ia) uniformly calibrated low-$z$ dataset in the \textit{southern sky} to date \citep{Sherman_25_DEBASSRelease}. They have already released 77 spectroscopically confirmed SNe~Ia that pass cosmology cuts in DR 0.5 \citep{acevedo2025darkenergybedrockallsky}. DEBASS operates in the southern sky with a similar redshift range as TITAN (0.01 $<$ z $<$ 0.08) and here we utilize $>400$ SNe~Ia in DEBASS DR1 (Sherman et al. in prep) which results in 63 matches with TITAN cosmology-quality light curves.  \rev{DEBASS claim high signal to noise, low Hubble residual scatter ($0.1$ mag) light curves, resulting in a reasonably strong constraint on the offset between ATLAS and DEBASS.} This should enable excellent cross matching of SNe~Ia once DEBASS DR1 is released and we can find hundreds of matches. \rev{Becasue DEBASS is calibrated to DES Y6, and, as demonstrated in this paper, ATLAS is now tied to DES Y6 as well, we do not use any offsets here.} 

\subsection{YSE}

YSE (Young Supernova Experiment) is comprised of data from ZTF and PS1, and contains 451 spectroscopically confirmed and cosmology grade SN Ia light curves \citep{Aleo_23_YSERelease}. \revtwo{We find 35 cross matches with TITAN.} The YSE redshift range is generally higher than TITAN ($z < 0.5$), \rev{yet this still results in 35 matches with our TITAN low-$z$ dataset.} \rev{We use the PS1-Dovekie \citep{Popovic_2025_Dovekie} offsets relative to DES-Dovekie in order to place YSE on the DES Y6 system and to facilitate comparison with TITAN.} 

\subsection{Comparison of $\mu_\mathrm{test}$ for Coincident SNe~Ia}

\rev{In Fig.~\ref{fig:titan_survey_mu}, the observed $\mu_{test}$ for coincident SNe~Ia between TITAN and both DEBASS and YSE is found to be in agreement, with average offsets between the surveys of $-0.005\pm0.012$ and $+0.039\pm0.025$ respectively. We do find, as expected, a significant offset for ZTF DR2 ($-0.038\pm0.007$) suggesting that the ZTF DR2 photometry is bright relative to TITAN. While this ZTF DR2 offset is in agreement with the offset presented in \cite{Newman_2025_redgiant}, we do not find strong evidence for an offset of ZTF DR2 at the 90mmag level as presented in \cite{lacroix2025ztfsneiadr2}.} \rev{Overall, in comparison to the the surveys that have been used in modern cosmology analyses (YSE/PS1 and DEBASS/DES) we find no direct evidence of systematics.}

\section{Discussion} \label{sec:discussion}

\begin{deluxetable}{ccccc}
    \vspace{0.3cm}
    \tablecaption{\rev{Slopes and offsets equivalent to Fig.~\ref{fig:primary_calib} for all chips and filters. All values displayed here are post inter--chip and wave shift correction. These values are used in our second two systematics in Tab.~\ref{tab:systematics}.}}
    \label{tab:primary_slope_offsets}
    \tablehead{
        \colhead{Chip} & 
        \colhead{$\Delta m_{primary}(\textit{o})$ (mag) } & 
        \colhead{Primary Slope (\textit{o})} & 
        \colhead{$\Delta m_{primary}(\textit{c})$ (mag) } & 
        \colhead{Primary Slope (\textit{c})} 
    }
    \startdata
    0 & $-0.003 \pm 0.014$ & $+0.001 \pm 0.005$ & ... & ... \\
    1 & $+0.002 \pm 0.009$ & $+0.001 \pm 0.005$ & $-0.012 \pm 0.014$ & $-0.013 \pm 0.005$ \\
    2 & $+0.003 \pm 0.015$ & $+0.011 \pm 0.005$ & ... & ... \\
    3 & $-0.001 \pm 0.011$ & $+0.007 \pm 0.006$ & $-0.003 \pm 0.022$ & $-0.005 \pm 0.007$ \\
    4 & $+0.010 \pm 0.011$ & $+0.001 \pm 0.007$ & $-0.011 \pm 0.016$ & $-0.009 \pm 0.005$ \\
    5 & $+0.005 \pm 0.010$ & $+0.002 \pm 0.006$ & $-0.004 \pm 0.012$ & $-0.009 \pm 0.005$ \\
    6 & $-0.002 \pm 0.011$ & $+0.005 \pm 0.004$ & $-0.006 \pm 0.013$ & $-0.000 \pm 0.004$ \\
    7 & $-0.001 \pm 0.012$ & $-0.007 \pm 0.009$ & $+0.000 \pm 0.015$ & $-0.002 \pm 0.010$ \\
    8 & $+0.023 \pm 0.013$ & $-0.011 \pm 0.011$ & $+0.008 \pm 0.020$ & $-0.003 \pm 0.016$ \\
    \enddata

    \vspace{-0.8cm}
\end{deluxetable}

    \subsection{Chromatic Effects} \label{sec:chromatic_effect}
    
   \rev{We identify a color-dependent calibration residual with an amplitude of approximately 0.005 to 0.045 mag over the $g-i$ range relevant for TITAN SNe Ia. The effect is most pronounced in the cyan filter, particularly for the sitecam 02a system (chips 4c, 5c, and 6c), as shown in Fig.~\ref{fig:slope_whiskers}. For chips 4c, 5c, and 6c there is a clear progression towards worsening color-dependence over time. Conversely, chips 7c and 8c, which are associated with the newer southern telescopes, have substantially smaller chromatic trends. This temporal and instrumental coherence points to an instrument-level effect. This could be explained plausibly by small mismatches between the assumed and true filter transmission functions, evolution of detector quantum efficiency, or wavelength-dependent throughput changes elsewhere in the optical system all of which could lead to color-dependent zero-point offsets. Resolving the physical cause of this chromatic effect will require further experimentation. The main thrust of the work presented here, is that we are able to adequately correct this chromatic effect for cosmology by shifting the filter throughput in wavelength.}

    \rev{\subsection{Intra--chip Correction for Use in Cosmology}}

\revthree{The intra–chip corrections derived in this work are generally small and spatially smooth for all detector configurations, with the exception of chip 8 in the orange band (8o). Chip 8o exhibits a pronounced spatial structure, with a clear gradient toward the right side and upper portion of the detector. While we construct and apply a correction map for this chip and include it in preliminary distance measurements, the amplitude and coherence of this feature distinguish it from the lower level intra–chip structure seen elsewhere in the focal plane.

Upon further investigation, we find no corresponding offset in the image-level zeropoint solution nor a spatially coherent deviation in the DOPHOT-based stellar photometry, indicating that the effect arises downstream in the forced photometry process. In particular, the discrepancy appears only in the forced photometry which utilizes a spatially variable PSF model. While this rules out a bug in the core ATLAS processing and implies peculiarities of the 8o PSF, the underlying cause is thus far unidentified. 

Given this, chip 8o represents the dominant contributor to residual intra–chip uncertainty in the current TITAN calibration. We do not attempt to further absorb this effect through ad hoc error inflation. Instead, our preferred approach is to identify and correct the underlying cause of the forced photometry behavior prior to cosmological analyses. Our intra-chip correction is applied in this DR1 release as it reflects the photometry exactly as produced by the current ATLAS pipeline used for cosmology-quality SN~Ia light curves being released by TITAN. Any upstream changes to PSF handling would result in future releases that will supersede this work. In future cosmological analyses we will evaluate the possible systematics due to the anomaly in chip 8o and the state of its potential resolution or lack thereof. }

    \subsection{Systematic Uncertainty} \label{sec:systematics}

\begin{deluxetable*}{lccccl}
    \tablecaption{Average systematic uncertainty values per filter before and after calibration. Values are in magnitude. \label{tab:systematics}}
    \tabletypesize{\footnotesize}
    \tablehead{
     & \multicolumn{2}{c}{orange} & \multicolumn{2}{c}{cyan} & \\
    \colhead{\textbf{Systematic}} &
    \colhead{Before} &
    \colhead{After} &
    \colhead{Before} &
    \colhead{After} &
    \colhead{Description\tablenotemark{a}}
    }
    \startdata
    Intra--chip (pixel-to-pixel) variation & 0.007 & 0.003 & 0.005 & 0.003 & $\sigma(\Delta\mathrm{ZP}_\mathrm{pixel})$ in Fig.~\ref{fig:intrachip_allchip}\\
    Inter--chip (chip-to-chip)  variation & 0.017 & 0.002 & 0.016 & 0.003 & $\sigma(\langle\Delta\mathrm{ZP}\rangle_\mathrm{chip})$ in Fig.~\ref{fig:offset_validation} \\
    Chromatic Effect & 0.005 & 0.004 & 0.029 & 0.005 & Median slope ($\bar{\mathcal{A}}$) $\times$ SNe~Ia color range (Sec.~\ref{sec:systematics}, \rev{Tab.~\ref{tab:primary_slope_offsets}})\\
    Absolute Calibration & 0.012 & 0.003 & 0.007 & 0.006 & Median size of CALSPEC offsets, \rev{median $\Delta$m in Tab.~\ref{tab:primary_slope_offsets}} \\
    \hline
    \textbf{Total} & \textbf{0.022} & \textbf{0.005} & \textbf{0.034} & \textbf{0.009} \\ 
    \enddata
    \tablenotetext{a}{Figures are cited here for reference purpose only, as they may only show measurements made before or after correction. We measure the same quantity before and after applying our correction models to quantify the reported values in this table.}
\end{deluxetable*}

    In this paper we present \rev{preliminary estimates of systematic uncertainties due to calibration for future TITAN cosmology constraints. We define 4 sources of systematic uncertainty in this work and they are summarized in Table~\ref{tab:systematics}.} The first is the systematic uncertainty on the intra--chip correction, \rev{resulting in a per-exposure magnitude error floor}. This is calculated as: $\epsilon_{intra} = \sigma(\mathcal{O - M})$, where $\mathcal{O}$ is the offset across all pixels in the chip and $\mathcal{M}$ is the median of the chip (Sec.~\ref{sec:intrachip}). To determine this systematic post calibration we subtract out our correction map from the real data observed ($=\mathcal{O}-$correction-map) in Fig.~\ref{fig:intrachip_allchip} and recalculate the standard deviation of the offset per pixel. In Tab.~\ref{tab:systematics}, we see a $\sim3$ mmag improvement across both bands, which is substantial given that the initial effect is only $\sim7$ mmag. 
    
    A second magnitude error floor comes from the residual systematics on our inter--chip correction (Sec.~\ref{sec:inter-chip}). We find this by taking the standard deviation of the values presented in Fig.~\ref{fig:offset_validation} for each filter. Tab.~\ref{tab:systematics} shows that our systematic uncertainty in error floor improves in both orange and cyan (by $15$ and $13$ mmag respectively) after employing the inter--chip corrections.

    Next, we quantify the systematic uncertainty related to the \rev{chromatic wavelength shifts applied to ATLAS passbands and validated by our HST CALSPEC and DAWD validation sets} (Sec.~\ref{sec:inter-chip}, ~\ref{sec:chromatic_effect}). We define this systematic uncertainty as: $\epsilon_{chromatic} = \mathcal{\bar{A} *SN}$, where $\bar{\mathcal{A}}$ is the median slope across all chips, and $\mathcal{SN}$ is the observed SN~Ia color range. This is practically propagated from the observed slope in the HST CALSPEC residuals using the SNe~Ia color distribution ($-0.97\le g-i \le -0.11$ mag at 2-$\sigma$ tails, covering 95\% of the dataset) measured in \cite{Murakami_TITAN_DR1}. 
    Before our corrections, the median slopes for each filter is $\frac{\mathrm{Residual}}{g-i}\sim 0.029, 0.004$ for cyan\footnote{The slope for cyan band varies by a factor of a few between detectors. We use the median values for each filter as a representative value solely for the comparison with the post-correction size.} and orange band, respectively (see Fig.~\ref{fig:slope_whiskers}). This corresponds to $\sim0.026$ and $\sim 0.003$ mag-level changes in the zeropoint across the color range of SNe~Ia. The slope is consistent across our tertiary catalog and the primary, CALSPEC validation set.
    After applying our corrections derived from the tertiary star catalog, the remaining slope in the CALSPEC stars become considerably small (0.005), making the systematic uncertainty (See Table~\ref{tab:systematics}) consistent across filters $0.005\frac{\mathrm{mag}}{g-i\ \mathrm{mag}} \times (-0.11-(-0.97))\ \mathrm{mag} \approx 0.0043$ mag.

    Finally, we quantify our confidence in the absolute calibration of ATLAS using HST CALSPEC stars. We take the fitted offset (intercept of the slope at g - i color = 0) after the wave shift has been calculated ($\mathcal{B}$). The systematic before calibration then: $= \mathrm{Median(|\mathcal{B}|)} - \delta_m$. After calibration: = $= \mathrm{Median(|\mathcal{B}|) - (interchip)} - \delta_m$. This acts as an independent validation of only our inter--chip correction using primary calibrators. We specifically use data post-wave-shift correction for both the pre, and post absolute calibration systematic, as the wave-shift systematic is already contained within row 3 of Tab.~\ref{tab:systematics} (Sec.~\ref{sec:systematics}, Tab.~\ref{tab:primary_slope_offsets}). We find a 9 mmag, and 1 mmag improvement for orange and cyan bands respectively. 

    Table~\ref{tab:systematics} demonstrates that before our \rev{inter--chip, intra--chip, and wavelength} corrections, there exists a total systematic uncertainty of 22mmag and 35mmag in the orange and cyan bands respectively. After calibration we are able to reduce this to 5mmag and 10mmag respectively. For reasons discussed in \cite{Murakami_TITAN_DR1}, we find that in \textit{SNANA} we must add a 10 mmag error floor to our TITAN SNe~Ia already. This implies that our additional systematics from calibration are on a scale \rev{that do not significantly impact TITAN prospects for SN~Ia cosmology. }

\begin{figure*}
    \centering
    \includegraphics[width=\linewidth]{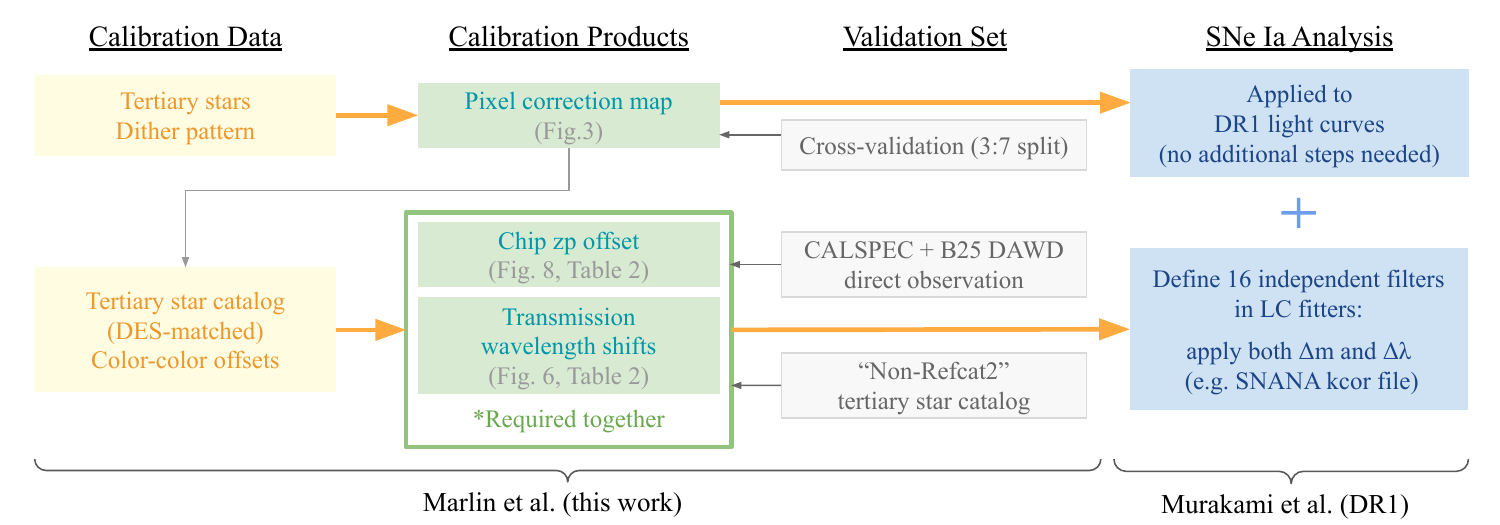}
    \caption{A summary of this work, products, and usage in future analysis. From left to right: dataset used in our analysis, calibration products (optimal calibration), dataset used to validate our calibration, and the methods to apply our calibration to SNe~Ia light curve analysis (e.g., SALT3 fitting).}
    \label{fig:flowchart}
\end{figure*}

    \subsection{Usage and Data Tools}

    The substantial work presented here can be reduced to a simple calibration pipeline shown in Fig.~\ref{fig:flowchart}. This flow chart shows our three calibration outputs, the pixel correction map (intra--chip), chip to chip ZP offsets (inter--chip) and transmission wavelength shifts. For the chip ZP offsets and the transmission wavelength shifts, these values can be lifted directly from Tab.~\ref{tab:offset_comparison} and applied to any ATLAS photometry files using the tools presented in the \textit{ATLAST} \citep{atlast_repo} package of the data release. The pixel correction map produced here will also be available for download with DR1, and can be applied with a single line of python code from \textit{ATLAST}. 

    Fig.~\ref{fig:flowchart} also shows the validation sets used. The chip ZP offset and transmission wavelength shift validations are discussed further in Sec.~\ref{sec:validation_with_calspec} and Sec.~\ref{sec:validation_independent_tertiary_catalog}, while the pixel correction map validation map is presented in App.~\ref{appendix: optimal_smoothing_radius}. For direct application to SN~Ia light curves please see \cite{Murakami_TITAN_DR1}. All tools and calibration data can be downloaded from: \url{https://titan-snia.github.io}.

\section{Conclusion} \label{sec:conclusion}

SNe~Ia are a well proven tool for measuring relative distances for use in cosmology. Until now, most major SN~Ia cosmology surveys have relied on the same 200 low-$z$ SNe~Ia. TITAN now provides the largest, independent, spectroscopically-confirmed low-$z$ SNe~Ia dataset to date. TITAN, which uses the ATLAS all-sky survey \rev{and data reduction pipelines \citep{ATLAS,atlasserver},} must be internally and externally calibrated before it can be used for cosmology. That calibration has been presented in this paper. 

We conduct a relative calibration between ATLAS and DES, as DES is a well-measured \rev{southern-sky survey that contains stars} both inside and outside of PS1 (the primary calibrating instrument of ATLAS Refcat2). We produced three distinct tertiary calibration star catalogs (Sec.~\ref{sec:tertiary_star_samples}) \rev{from the DES Y6 dataset:} 1) a `color--blind' color-uniform sample from across the DES footprint matched to Refcat2 stars, 2) a `blue' sample that is intentionally biased substantially blue to match the colors of low-$z$ SNe~Ia, and 3) a 'non-Refcat2' sample of stars that exist in DES but that do not exist within Refcat2 which provides a completely independent dataset that mimics the behavior of SNe~Ia in ATLAS. For each of these catalogs we request ATLAS photometry from the server. We also generate synthetic data from HST CALSPEC, DAWD, and NGSL specgra.

We examine pixel-to-pixel variations \textit{within} each ATLAS CCD and filter (`intra--chip') to build correction maps for each. We find most exhibit modestly small pixel-level structures below the 0.01 mag level, with the exception of chip 8o, where we notice a significant vignetting pattern (Fig.~\ref{fig:intrachip_allchip}). To account for the variations in pixel we build a correction map. This is produced by binning the data, smoothing at an optimally calculated pixel radius (see App.~\ref{appendix: optimal_smoothing_radius}), then remapping to the 10,560 x 10,560 pixel CCD. We separately produce a map for each chip-filter combination. 

We also compute corrections \textit{across} each CCD and filter (`inter--chip'). We define this as the vertical offset in DES - ATLAS transformation and stellar color between the observed data and the synthetically produced data from NGSL (Fig.~\ref{fig:offset_plot_example}), following the likelihood defined in Sec.~\ref{appendix:multi-color_joint_likelihood}. Notably, the synthetic NGSL - real ATLAS residuals exhibit significant slopes in ATLAS cyan bands (Fig.~\ref{fig:slope_whiskers}). Following \cite{Popovic_2025_Dovekie}, we correct for this by applying a shift in the wavelengths of the filters (Fig.~\ref{fig:filter_shift}). 

 We validate our corrections in three ways: 1) with the independent 'non-Refcat2' tertiary star catalog (Fig.~\ref{fig:offset_validation}), 2) with independent primary and secondary absolute calibrators HST CALSPEC, and DAWD stars (Fig.~\ref{fig:primary_calib}), and 3) by comparing distance moduli of cross-matched SNe~Ia (Fig.~\ref{fig:titan_survey_mu}). All validation efforts point to improved consistency overall and reduced systematics (Tab.~\ref{tab:systematics}).

The calibration presented here serves as a \rev{baseline calibration, validation, and calibration-related systematic error budget} for the upcoming TITAN DR1 cosmological analysis. The data release and all associated tools will be presented on the TITAN website at: \url{https://titan-snia.github.io}. The light curves, host galaxies, and simulations will be presented in \cite{Murakami_TITAN_DR1}, \cite{Tweddle_2026_hosts}, and \cite{Tweddle_2026_simulation} respectively.

\begin{acknowledgements}

This work has made use of data from the Asteroid Terrestrial-impact Last Alert System (ATLAS) project. The Asteroid Terrestrial-impact Last Alert System (ATLAS) project is primarily funded to search for near earth asteroids through NASA grants NN12AR55G, 80NSSC18K0284, and 80NSSC18K1575; byproducts of the NEO search include images and catalogs from the survey area. This work was partially funded by Kepler/K2 grant J1944/80NSSC19K0112 and HST GO-15889, and STFC grants ST/T000198/1 and ST/S006109/1. The ATLAS science products have been made possible through the contributions of the University of Hawaii Institute for Astronomy, the Queen’s University Belfast, the Space Telescope Science Institute, the South African Astronomical Observatory, and The Millennium Institute of Astrophysics (MAS), Chile, and the University of Oxford. 

We would like to thank John Tonry and the ATLAS team for assistance in gathering massive amounts of forced photometry and for their conversations and investigations into the difference imaging pipeline and nuances of the telescopes (especially chip 8o). 

We thank the Hariri Institute at Boston University for their generous funding contributions to this project and for facilitating the space to hold a summer 2025 collaboration meeting. 

D.O.J. acknowledges support from NSF grants AST-2407632, AST-2429450, and AST-2510993, NASA grant 80NSSC24M0023, and HST/JWST grants HST-GO-17128.028 and JWST-GO-05324.031, awarded by the Space Telescope Science Institute (STScI), which is operated by the Association of Universities for Research in Astronomy, Inc., for NASA, under contract NAS5-26555.

SJS and KWS acknowledge funding from STFC Grant ST/Y001605/1, a Royal Society Research Professorship and the Hintze Family Charitable Foundation. 
\end{acknowledgements}

\appendix

\section{Multi-color joint likelihood analysis} \label{appendix:multi-color_joint_likelihood}

In Sec.~\ref{sec:inter-chip}, we fit a single offset value to an ATLAS filter so that the empirical ATLAS-DES filter transformation matches synthetic prediction. This transformation is color-dependent, and there are multiple possible combinations of DES filters (e.g., \texttt{ATLAS-o}$\rightarrow$\texttt{DES-g} as a function of color \texttt{DES-g}-\texttt{DES-i}). Each of the combination can be simultaneously evaluated to form a joint likelihood, and we describe the formalism of our likelihood function and the process to prepare necessary quantities below.

\begin{deluxetable}{cllll|llll}
    \tablecaption{Combinations of filters used for the offset analysis. \label{tab:filter_comb}}
    \tablehead{
      \colhead{} & \multicolumn{4}{c}{ATLAS-$c$} & \multicolumn{4}{c}{ATLAS-$o$}\\
      \colhead{\#} & \colhead{$y1$} & \colhead{$y2$} & \colhead{$x1$} & \colhead{$x2$} &
      \colhead{$y1$} & \colhead{$y2$} & \colhead{$x1$} & \colhead{$x2$}
    }
    \startdata
    1 & ATLAS-$c$ & DES-$g$ & DES-$g$ & DES-$r$ & ATLAS-$o$ & DES-$r$ & DES-$g$ & DES-$r$ \\
    2 & ATLAS-$c$ & DES-$g$ & DES-$r$ & DES-$i$ & ATLAS-$o$ & DES-$r$ & DES-$r$ & DES-$i$ \\
    3 & ATLAS-$c$ & DES-$g$ & DES-$g$ & DES-$i$ & ATLAS-$o$ & DES-$r$ & DES-$g$ & DES-$i$ \\
    4 & ATLAS-$c$ & DES-$r$ & DES-$r$ & DES-$i$ & ATLAS-$o$ & DES-$i$ & DES-$g$ & DES-$r$ \\
    5 & ATLAS-$c$ & DES-$r$ & DES-$g$ & DES-$i$ & ATLAS-$o$ & DES-$i$ & DES-$g$ & DES-$i$ \\
    \enddata
\end{deluxetable}
\onecolumngrid

First, assuming that the DES filters and their star catalog values are well-calibrated, we obtain an ATLAS offset for each (i-th) star as the following:
\begin{equation}\label{eq:offset_def}
    \Delta_\text{i, x1, x2, y1, y2} = m_{i,y1}^\text{ATLAS} - m_{i,y2}^\text{DES} - f_{y1\rightarrow y2}^\text{synth}\left(m_{i,x1}^\text{DES} - m_{i,x2}^\text{DES}\right)\ ,
\end{equation}
where $m$ represent observed magnitudes of stars, with subscripts $y1$ for the ATLAS band of interest and $x1,\ x2,\ y2$ for DES bands we use as a reference. Considering the overlaps of the sensitivity functions, we use the combinations of filters shown in Table~\ref{tab:filter_comb}.
We note that there are two exceptions in the listed combinations: the dataset obtained with ($y1,\ y2,\ x1,\ x2$) = ($c$,$g$,$g$,$r$) is linearly identical to ($c$,$r$,$g$,$r$), and it causes the covariance matrix we describe later to be nearly singular. To avoid this issue and considering that it adds nearly no information, we exclude such combination. Similarly, another combination for the orange filter ($o$,$i$,$r$,$i$) is excluded.
The synthetic transformation function between ATLAS filter $y1$ and DES filter $y2$ $f_{y1\rightarrow y2}^\text{synth}$ is obtained by fitting a third-order polynomial to a fully synthetic data $m'$,
\begin{equation}
y\overset{\text{fit}}{=} f_\mathrm{synth} = \operatorname{Poly}_3(x),\quad
x = m'_{x_1}-m'_{x_2},\quad y = m'_{y_1}-m'_{y_2}\ .
\end{equation}

After evaluating $\Delta_i$ for each of the combination, we obtain a vector of offsets $\bf{r}_i = (\Delta_{i1},\ \Delta_{i2},\ \ldots, \Delta_{i5})^\top$. Due to repeated uses of data, these measurements are not independent from each other, and we quantify that effect by constructing a filter-to-filter covariance matrix for each star.
Propagating uncertainties from each observed quantity in Eq.~\ref{eq:offset_def}, we obtain a $5\times5$--matrix:
\begin{align}
[\Sigma_i]_{jk} &= 
\underbrace{ \sigma^2_{y_{1j}} + \sigma^2_{y_{2j}}\,\delta(y_{2j},\,y_{2k}) }_{\text{ATLAS $y_1$ - DES $y_2$ error}} \notag \\
&\quad + \underbrace{
    f'_{k} \sigma^2_{y_{2j}} \left[ \delta(y_{2j},\,x_{1k}) - \delta(y_{2j},\, x_{2k})\right]
    + f'_{j} \sigma^2_{y_{2k}} \left[ \delta(y_{2k},\,x_{1j}) - \delta(y_{2k},\,x_{2j}) \right]
}_{\text{$y2$ --- color}} \notag \\
&\quad + \underbrace{
    f'_{j} f'_{k} \sigma^2_{x_{1j}} \left[ \delta(x_{1j},\,x_{1k}) - \delta(x_{1j},x_{2k}) \right]
    - f'_{j} f'_{k} \sigma^2_{x_{2j}} \left[ \delta(x_{2j},\,x_{1k}) - \delta(x_{2j},\,x_{2k}) \right]
}_{\text{color --- color}} .
\end{align}

Using this covariance matrix, we obtain an appropriate weights between each measurement within $\bf{r}_i$ and collapse it into a single, representative offset value per star (generalized least-square estimation; GLS):
\begin{equation}
\bar r_i \;=\; \frac{\mathbf{1}^\top \Sigma_i^{-1} \mathbf{r_i}}{\mathbf{1}^\top \Sigma_i^{-1} \mathbf{1}},
\qquad
\sigma^2_{\bar r,i} \;=\; \frac{1}{\mathbf{1}^\top \Sigma_i^{-1} \mathbf{1}}\
\end{equation}
where $\sigma^2_{\bar r,i}$ is the variance for $\bar r_i$, and $\mathbf{1} = (1,1,\ldots, 1)^\top$ is an all-one vector.

The obtained per-star offset value $\bar r_i$ and its variance $\sigma^2_{\bar r,i}$ is then used to evaluate our likelihood, which accounts for possible combinations of filters, their uncertainties, covariances, and overlapping use of data across such combinations:
\begin{equation} \label{eq:residual_likelihood}
\ell(\Delta m_\text{f},\sigma_{\rm int, f})
= \sum_i^{N_\text{star}} -\frac{1}{2}\left[
\frac{(\bar r_i-\Delta m_\text{f})^2}{\sigma^2_{\bar r,i}+\sigma_{\rm int}^2}
+\ln\!\bigl(2\pi \sigma^2_{\bar r,i}+2\pi \sigma_{\rm int}^2\bigr)
\right]\ .
\end{equation}
This formula evaluates the likelihood of proposed offset for the ATLAS filter $\Delta_\text{f}$ (mag) against the par-star residual $\bar r_i$ (mag) for each $i$-th star, which is derived from multiple combinations of filters between ATLAS and DES. We simultaneously measure the star-to-star intrinsic scatter $\sigma_\text{int}$.

    \begin{figure*}
    \centering
    \includegraphics[width=\linewidth]{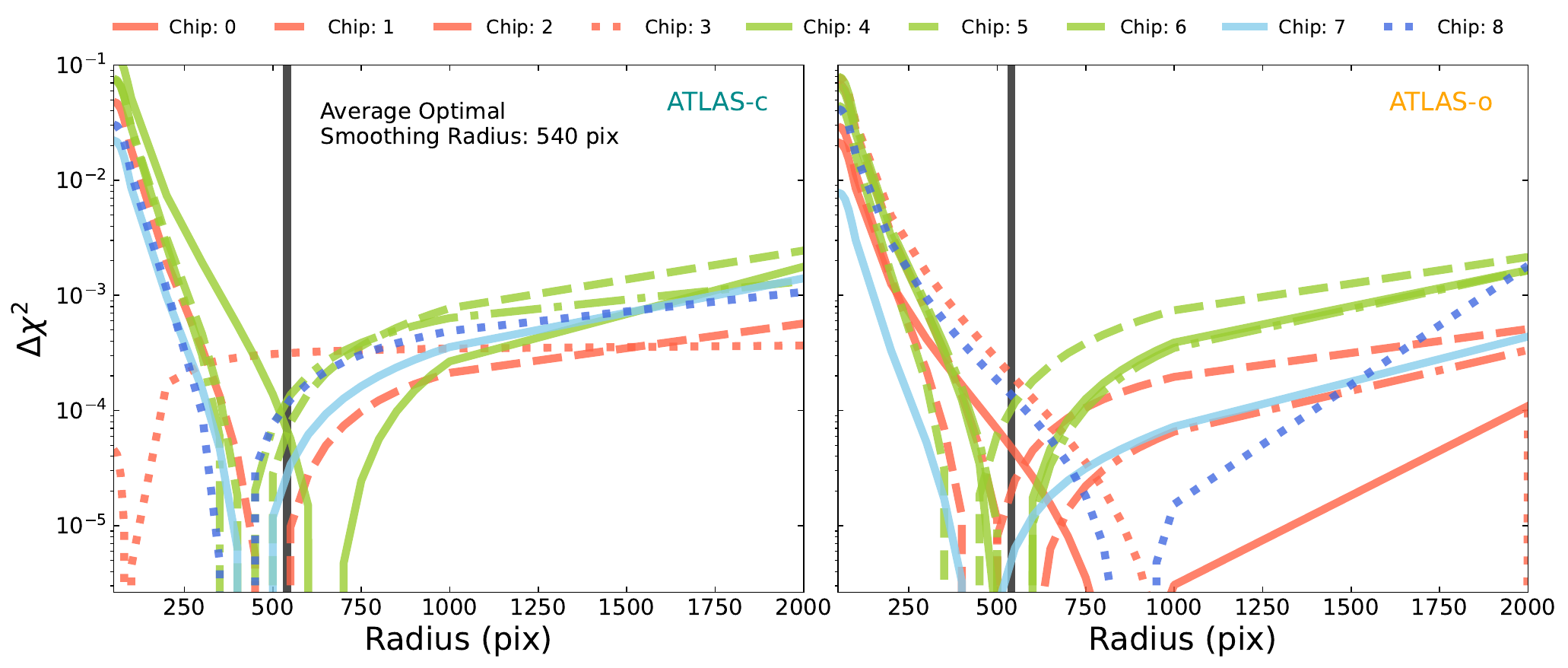}
    \caption{Cross validation plot of reduced $\chi^2$ vs smoothing radius in pixels. You can see the smoothing radius that minimizes the $\chi^2$ is focused around 540 pixels. We would rather slightly over bin (thus under correct) than under bin which would result in over correcting leading to potentially misleading and unrealistic trends. The y axis is reduced $\chi^2$ minus the minimum chi squared for each chip filter combo.}
    \label{fig:cross_validation_plot}
    \end{figure*}

\section{Profile likelihood for wavelength shift}
    We estimate the optimal wavelength shift in the transmission functions (Fig.~\ref{fig:filter_shift}) for each chip-filter combination using the profile likelihood method. 
    When we allow filter transmission function to have a small shift in the wavelength (which effectively changes the pivot wavelength and introduces/corrects the chromatic effect as described in Sec.~\ref{sec:inter-chip}), the color-averaged residual $\bar r_i$ in the likelihood function (Eq.~\ref{eq:residual_likelihood}) becomes a function of wavelength-shift size $\Delta\lambda_\text{filt}$. The updated log-likelihood is therefore
    \begin{equation}
        \ell(\Delta\lambda_\text{f},\Delta m_\text{f},\sigma_{\text{int},f}) = \sum_i^{N_\text{star}} -\frac{1}{2}\left[
\frac{\left[\bar r_i(\Delta\lambda_\text{f})-\Delta m_\text{f}\right]^2}{\sigma^2_{\bar r,i}+\sigma_{\rm int}^2}
+\ln\!\bigl(2\pi \sigma^2_{\bar r,i}+2\pi \sigma_{\rm int}^2\bigr)
\right]\ 
    \end{equation}
    and this is a computationally expensive as each likelihood call requires the synthetic photometry of CALSPEC and DAWD stars to be calculated with updated filter functions. Outlier rejection is often necessary to account for poor observing conditions or poor psf fit due to large proper motions, and varying data vector $\bar r_i$ makes it more difficult to retain a fixed, reproducible set of data while effectively rejecting outliers if one chooses to simply optimize all free parameters at once.
    We simplify this problem by evaluating profile likelihood along the $\Delta\lambda_\text{f}$ space,
    \begin{equation}
        \ln P(\Delta\lambda_\mathrm{f}) = \ell(\Delta\lambda_\text{f}, \widehat\Delta  m_\text{f},\widehat\sigma_{\text{int},f}) - \ell_\mathrm{max}\, ,
    \end{equation}
    where $\ell(\Delta\lambda_\text{f}, \widehat\Delta  m_\text{f},\widehat\sigma_{\text{int},f})$ is the likelihood maximized with a fixed $\Delta\lambda_\mathrm{f}$. 
    Practically, this is evaluated over a grid of $\Delta\lambda_\mathrm{f}$ between $-150 \le \Delta\lambda_\mathrm{f} \le 150$ \AA\xspace with 1\AA-spacing. The profile is then iteratively improved by applying an outlier rejection based on the maximum-likelihood set of $\bar r_i$ and using the same set of outlier-rejected stars across the grid.
    Once convergence of the profile is achieved, we fit a quadratic function to $\ln P(\Delta \lambda_f)$ to determine the best-fit offset $\Delta \lambda_\text{f,best}$ and its uncertainty $\sigma_\lambda$, assuming Gaussian posterior. The typical size of the uncertainty is $\sigma_\lambda\lesssim10$\AA, and it corresponds to $\sim$mmag level of systematic, which is included in our analysis but is negligible compared to the estimated size of the systematic uncertainty from validation set.

\section{Optimal Smoothing Radius Determination:} \label{appendix: optimal_smoothing_radius}

    To determine our optimal smoothing radius for the intra--chip correction, we trained a model on 70\% of the data from our calibration stars and then validated it with the remaining 30\%. This process is done on each chip/ filter combo and the split is regenerated randomly four times for each combo. This results in Fig.~\ref{fig:cross_validation_plot}, which shows the reduced $\chi^2$ as a function of smoothing radius. You can see that most of the chips are in the 250 - 750 pixel smoothing radius range for minimum reduced chi squared.

    Our model functions as follows: it bins the data into 50x50 pixel chunks. It then convolves the binned data with a gaussian kernel (it ignores edge effects as these have a higher likelihood of being inaccurate by definition). Our model then uses the large scale structure of the CCD to correct for systematic offsets in the photometric residuals across the chip.
    
    We used Fig.~\ref{fig:cross_validation_plot} to determine a median minimum $\chi^2$ across all chips, for which smoothing radius we should use for our correction model. We then apply the smoothing function to the dataset which provides a correction to the dataset specifically correcting the chip 8 data without changing the rest of the chips data in a non-uniform way. This produces our intra--chip correction.

\bibliographystyle{mn2e}
\bibliography{main}{}

\end{document}